\documentclass{aa}
\usepackage{graphicx}
\usepackage{txfonts}
\usepackage{subfigure}
\usepackage{multirow}
\usepackage{natbib}
\usepackage{appendix}
\usepackage{array}
\usepackage{soul}
\usepackage[normalem]{ulem}

\begin{document}
   \title{Multiwavelength campaign on Mrk 509}

   \subtitle{IX. The Galactic foreground}

   \author{C. Pinto
          \inst{1}
          \and
          G.~A.~Kriss \inst{2,3}
          \and
          J.~S.~Kaastra\inst{1,4}
          \and
          E.~Costantini\inst{1}
          \and
          J.~Ebrero\inst{1}
          \and
          K.~C.~Steenbrugge\inst{5,6}
          \and
          M.~Mehdipour\inst{7}
          \and
          G.~Ponti\inst{8}
          %\fnmsep\thanks{Just to show the usage
          %of the elements in the author field}
%           C.~P.~de~Vries\inst{1}
          }

% J. Ebrero1, G. A. Kriss2, J. S. Kaastra1,3, R. G. Detmers1,3, K. C. Steenbrugge4, E. Costantini1,
% N. Arav5, M. Cappi6, G. Branduardi-Raymont7 , M. Mehdipour7, P. O. Petrucci8, G. Ponti9, and C. Pinto1

   \institute{SRON Netherlands Institute for Space Research,
              Sorbonnelaan 2, 3584 CA Utrecht, The Netherlands\\
              \email{c.pinto@sron.nl}
         \and
              Space Telescope Science Institute, 3700 San Martin Drive, Baltimore,
              MD 21218, USA
         \and
              Department of Physics and Astronomy, The Johns Hopkins University, Baltimore, MD, 21218
         \and
             Astronomical Institute, Utrecht University,
             P.O. Box 80000, 3508 TA Utrecht, The Netherlands
         \and
             Instituto de Astronomia, Universidad Catolica del Norte, Avenida Angamos 0610, Antofagasta, Chile
         \and
             Department of Physics, University of Oxford, Keble Road, Oxford, OX1 3RH, UK
         \and
             Mullard Space Science Laboratory, University College London, Holmbury St Mary, Dorking, Surrey, RH5 6NT, UK
          \and
             School of Physics and Astronomy, University of Southampton, Highfield, Southampton, SO17 1BJ, UK\\
            }

   \date{Received December 19, 2011; accepted March 22, 2012}

% \abstract{}{}{}{}{} 
% 5 {} token are mandatory
 
  \abstract
   {The diffuse gas in and nearby the Milky Way plays an important role in the evolution of the entire Galaxy. It has a complex structure characterized by neutral, weakly and highly ionized gas, dust, and molecules.}
   {We probe this gas through the observation of its absorption lines in the high-energy spectra of background sources.}
   {We use high-quality spectra of AGN Mrk 509, located at high Galactic latitudes obtained with XMM-Newton, HST and FUSE. We use advanced absorption models consisting of photo- and collisional-ionization.}
   {We constrain the column density ratios of the different phases of the interstellar medium (ISM) and measure the abundances of C, N, O, Ne, Mg, Al, Si, S, and Fe. We detect seven discrete interstellar clouds with different velocities. One is a typical low-velocity cloud (LVC) and three belong to the family of the intermediate-velocity clouds (IVCs) found near the Galactic disk. These four clouds show large deviation from Solar abundances in the gas phase, mostly caused by dust depletion. The other three clouds are ionized high-velocity clouds (HVCs) and are located either in the Galactic environment or in the Local Group halo as suggested by the signatures of collisional ionization. The {similar abundances and ionization structure} of the HVCs suggest a common location and origin: they might belong to the remainder of an extragalactic cloud which was captured by the Galaxy.
}
   {We have shown that combined UV / X-ray spectroscopy is a powerful tool to investigate the ISM. In common Galactic clouds, like LVCs and IVCs, the ISM shows a complex structure consisting of at least three different temperature phases.
}
  %{Our work confirms that high resolution X-ray spectroscopy is a powerful tool to probe the interstellar medium. We carry out a complete diagnostic of the ISM: the chemical analysis, the study of the temperature structure and the different ionization states.}


   \maketitle
%
%________________________________________________________________

\section{Introduction}
\label{sec:introduction}

The interstellar medium (ISM) drives the evolution of the Galaxy: it is enriched with heavy elements during the course of stellar evolution, and it also provides the source of material for the subsequent star formation. In the spectra of background sources the ISM gives rise to reddening and absorption lines. The ISM shows a clear multi-phase structure \citep[for a review, see][]{ferriere}. The cold phase is a blend of dust, molecules and gas below $10^4$ K. The warm ionized gas is weakly ionized, with a temperature of $\sim 10^4$ K. The hot ionized gas is characterized by temperatures of about $10^{6}$ K.

The multi-phase medium plays an important role in the evolution of the Galaxy. One of the parameters that affect the stellar evolution is the metallicity of the star forming regions. Stellar winds and supernovae expel part of the interstellar gas out of the Galactic disk, but gravity generally stops the gas from escaping, such that it falls back through the process known as ''Galactic fountain'' \citep{Shapiro1976}. Gas accreted from smaller galaxies, e.g. the Magellanic Clouds, and the intergalactic medium increases the reservoir of low metallicity gas. High-velocity clouds (HVCs) play a crucial role in this process.

HVCs contain neutral hydrogen at velocities incompatible with a simple model of differential Galactic rotation. In practice one uses a Local Standard of Rest velocity $V_{\rm LSR} \geq 90\,{\rm km\,s}^{-1}$ to define HVCs \citep[for a review see][]{Wakker1997}. ISM absorbers with $30 \lesssim V_{\rm LSR} \lesssim 90\,{\rm km\,s}^{-1}$ are usually defined as intermediate-velocity clouds (IVCs). Both IVCs and HVCs might originate from our Galaxy or have an external origin. They could be debris from Galactic fountains or infalling Local Group gas \citep{Blitz1999}. Of course, a Galactic fountain origin would imply metallicities near solar, while infall from the Local Group or the intergalactic medium (IGM) would imply metal poor gas. The presence of infalling matter is required in order to maintain star formation in the Galaxy: without a substantial replenishing of the gas available the star formation would stop in a period much shorter than the Hubble Time. HVCs are thought to contain enough mass to sustain star formation rate (SFR) in the Galaxy \citep{Lehner2011}.

A multiwavelength approach provides the means to analyze the ISM in a complete way. For instance, combined UV and X-ray spectroscopy provides accurate column densities and the velocity structure of the most abundant ionic species. Ionic species like \ion{Si}{ii-iv}, \ion{C}{ii-iv}, \ion{N}{v} and \ion{O}{vi} especially, are quite common in HVCs \citep{Sembach2003}. In the last decade interstellar absorption lines in UV spectra of background stars have provided the distances of several IVCs and HVCs \citep[see e.g.][and references therein]{Richter2006, Lehner2011}. In summary, all the main \ion{H}{i} complexes are known to be Galactic. IVCs are at typical distances of $\lesssim 2$\,kpc, while HVCs with $90 \lesssim V_{\rm LSR} \lesssim 170\,{\rm km\,s}^{-1}$ are found between 4--13\,kpc. No cloud is found at $V_{\rm LSR} \geq 170\,{\rm km\,s}^{-1}$ towards halo stars, while several of them are found towards AGNs, which suggests that these very-high velocity clouds (VHVCs) are at larger distances. However all are thought to be within 40\,kpc \citep{Richter2006}, {except the Magellanic Stream which is at about 50\,kpc,} much closer than the typical distances in the Local Group halo. A plausible scenario is that VHVCs are the next generation of HVCs infalling towards the Galactic disk and slowing down during their interaction with it \citep{Lehner2011}.

The AGN \object{Mrk 509} has been intensively studied for both its intrinsic spectral features and its interesting line-of-sight. The Galactic latitude for Mrk~509 is --30 degrees and crosses the halo of our Galaxy, resulting in an important contribution of ionized gas. First evidence for ionized gas was found by \citet{York1982}, who discovered significant absorption from \ion{Si}{ii}, \ion{Fe}{ii} and \ion{C}{iv} in the spectra taken with the IUE satellite. They also observed strong Ly$\alpha$ absorption as well as red-shifted lines from \ion{Ca}{ii} and \ion{Na}{ii} in optical spectra. \citet{Blades1983} attributed these shifted lines to corotating gas in our halo. This was confirmed by the detection of neutral hydrogen with $V_{\rm LSR} \sim 60 \; {\rm km \; s}^{-1}$ \citep{McGee1986}. \citet{Sembach1995} and \citet{Sembach1999} discovered two sets of lines of \ion{Si}{iii-iv} and \ion{C}{iv} with large velocities, i.e. $V_{\rm LSR} \sim -230 \; {\rm km \; s}^{-1}$ and $-280 \; {\rm km \; s}^{-1}$, which were attributed to absorption by HVCs. \citet{Sembach2003} argued that these clouds are photo-ionized by the extra-Galactic background, but the presence in large quantities of hot gas such as \ion{O}{vi} cannot be explained by the photo-ionization models and suggests that these clouds are probably interacting with the hot Galactic corona or the Local Group medium. To explain the ionized C, Si and O column densities Collins et al. (2004) concluded that the HVCs have multiple phases. They showed that the \ion{Si}{iii-iv} and \ion{C}{iv} can be produced by a QSO photo-ionizing background, while the \ion{O}{vi} indicates collisional ionization due to the {interaction with the Galactic corona}. We note that the presence of molecular ${\rm H}_2$ in the LOS \citep{Wakker2006} affects the spectral region near the \ion{O}{vi} UV line and its column density estimate.

More than fifteen years after the discovery of these HVCs, their structure is not yet well understood. To solve these questions improved collisional and photo-ionization models are needed to constrain the ionization processes occurring in HVCs, as well as a multiwavelength approach which uses all the available archival data in order to increase the number of ions detected as well as the ionization parameter range that we can sample, and to disentangle the molecular H$_2$ and the \ion{O}{vi} absorption.

This article is one of a series of papers analyzing the deep and broad multiwavelength campaign on Mrk\,509. The overview of the campaign is presented in \citet{Kaastra11a}, hereafter paper\,I. Here we present the analysis of the interstellar clouds in the LOS towards Mrk 509 through a combined UV / X-ray analysis of the spectra taken with the \textit{Hubble Space Telescope} / Cosmic Origin Spectrograph (\textit{HST}/COS) \citep{Green2011} and the \textit{XMM-Newton} Reflection Grating Spectrometer \citep[RGS, ][]{denherder01}. Our analysis is focused on both the LVC, IVCs and HVCs and the dust present along the LOS towards Mrk\,509. We constrain velocity, ionization and chemical structure of the ISM for the different ionic species.

The paper is organized as follows. In Sect.~\ref{sec:data} we present the data. In Sect.~\ref{sec:spectra} we report the relevant spectral features that we analyze. In Sect.~\ref{sec:analysis} we describe the models we use and the results of our analysis. The discussion and the comparison with previous work are given in Sect.~\ref{sec:discussion}. Conclusions are reported in Sect.~\ref{sec:conclusion}.

\section{The data}
\label{sec:data}

As part of our multiwavelength campaign we observed Mrk\,509 in the far-ultraviolet
wavelength band between 1155\,\AA\ and 1760\,\AA\ using the Far-Ultraviolet
Channel and the medium resolution gratings of the Cosmic Origins
Spectrograph (COS) onboard the Hubble Space Telescope (HST). A detailed
explanation of the observing strategy, instrument performance, data
reduction, and calibration can be found in \citet{Kriss11b}, hereafter paper\,VI. They also present a full-scale plot of the high-resolution spectrum.

To summarize the data briefly, the observations were taken on 2009
December 10 and 11 simultaneously with the {\it Chandra} LETGS
observations in our campaign \citep{Ebrero11}. Using the COS gratings
G130M and G160M, we obtained total exposure times of 9470 s and 16452 s,
respectively. With each grating, we used only two different grating tilts
with a single FP-POS for each to avoid creating gaps in the spectral regions of interest including
the AGN outflow features.
The data were processed with the COS calibration pipeline v2.11b at STScI.
With central wavelengths of 1309 and 1327 for
G130M, 1577 and 1589 for G160M, and two exposures at each tilt, we
obtained a sufficient variety of independent placements of the spectrum
on the detector to identify instrumental features in our high signal-to-noise
spectrum. Known features such as dead spots were excluded from the
combined data; correctable features such as grid-wire shadows were
removed via a customized flat-field treatment described by \citet{Kriss11b}.
\citet{Kriss11b} also describe improved wavelength calibration
applied to the data. Finally, they deconvolved the
spectra using a Lucy-Richardson algorithm to remove the effects of the
broad wings of the COS line-spread function \citep{Ghavamian09, Kriss11a}.
This correction is essential to recover the true depths of narrow
interstellar absorption lines; comparison to prior STIS spectra of Mrk 509
validates the effectiveness of the deconvolution. We use the
original spectrum to identify weak spectral
features, while we use the deconvolved spectrum for measuring
the depth and width of identified absorption lines.

For the X-ray portion of our study we use the stacked {\it XMM-Newton}
RGS spectrum of Mrk 509 that consists of ten $\sim$60\,ks individual
observations. The observations and data reduction for this stacked RGS spectrum are
described by \citet{Kaastra11b}, hereafter paper\,II. Briefly, they used the SAS 9.0 software package to reduce the ten
individual observations. They then created a fluxed spectrum for
each observation and stacked these, using RGS 1 and 2 and both spectral
orders, and taking the effects of the {\it XMM-Newton} multi-pointing mode
into account.

In addition to the COS and RGS spectra, we also use an archival spectrum
obtained with the {\it Far Ultraviolet Spectroscopic Explorer} (FUSE) for
the \ion{O}{vi} $1032$\,{\AA} absorption line.
The FUSE data are described in paper\,VI.

For our spectral analysis we use SPEX\footnote{www.sron.nl/spex} version 2.03.00 \citep{kaastraspex}. SPEX is a software package optimized for the analysis and interpretation of high-resolution UV and X-ray spectra. In SPEX one can simultaneously use different components for continuum emission as well as absorption by photo-ionized or collisionally-ionized gas (see the SPEX manual). We scale elemental abundances to the proto-Solar values recently recommended by \citet{Lodders09}. We use a linear scale for the abundances. Throughout this paper we use SI units which are also the units system used in our spectral codes, adopt $1\sigma$ errors, and use $\chi^2$ statistics unless otherwise stated. We define high-ionization ions to be more than four times ionized, like \ion{O}{vi-viii}, while single to four times ionized ions are defined as low~/~intermediate-ionization ions, as commonly done in X-ray spectroscopy.

\section{Spectral features}
\label{sec:spectra}

\subsection{The COS spectrum}
\label{subsec:cos_spectrum}

\begin{figure*}
\begin{center}
      \subfigure{ 
      \includegraphics[width=12cm, angle=-90]{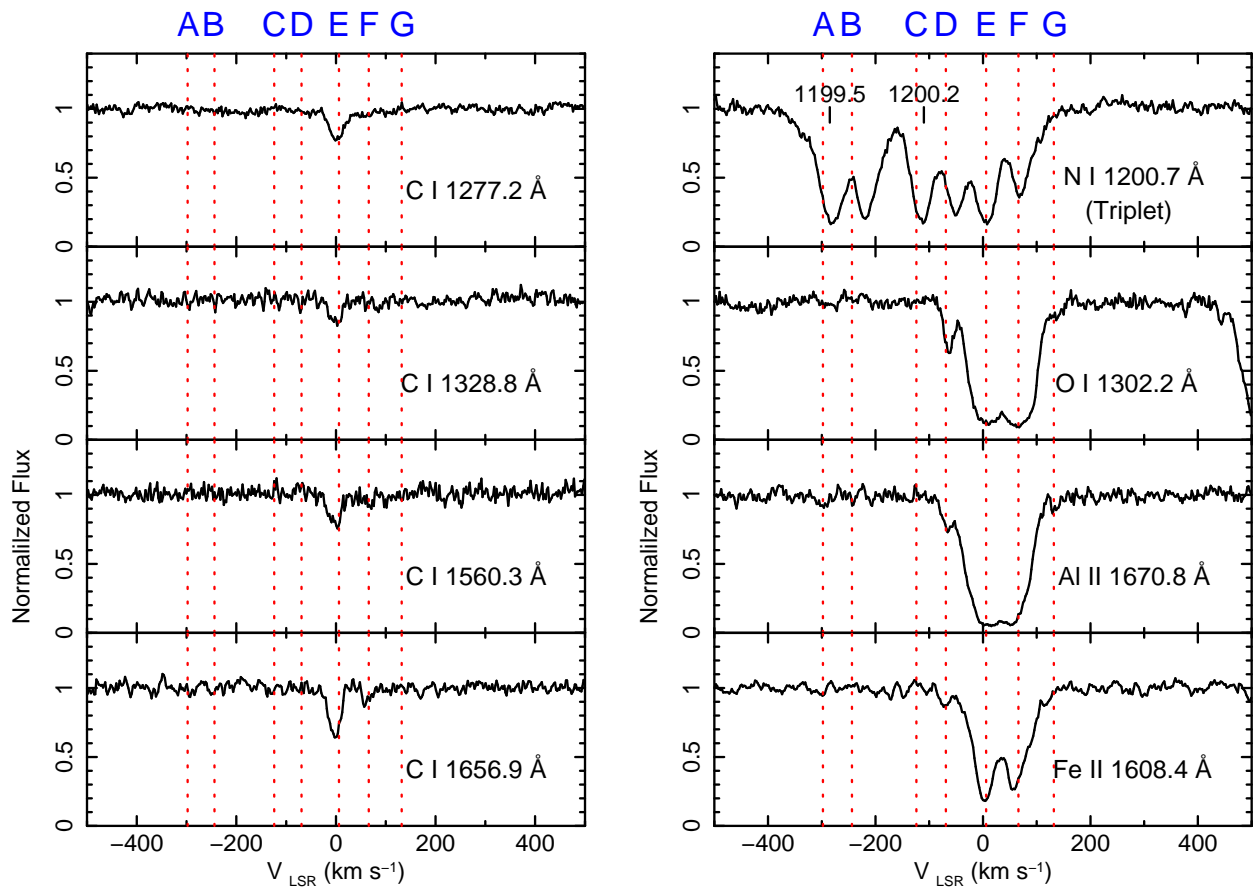}}
      \caption{HST/COS absorption lines from the cold phase {(ionization states I-II)}. The flux is normalized to the continuum emission and the lines are displayed in the Local Standard of Rest system. The rest frame wavelength of each resonance line is also given. The dotted lines represent the average velocities of the seven cloud systems \citep{Kriss11b}, which are also labeled as A~to~G.}
          \label{fig:cos_lines1}
   \end{center}
\end{figure*}

\begin{figure*}
  \begin{center}
      \subfigure{ 
      \includegraphics[width=12cm, angle=-90]{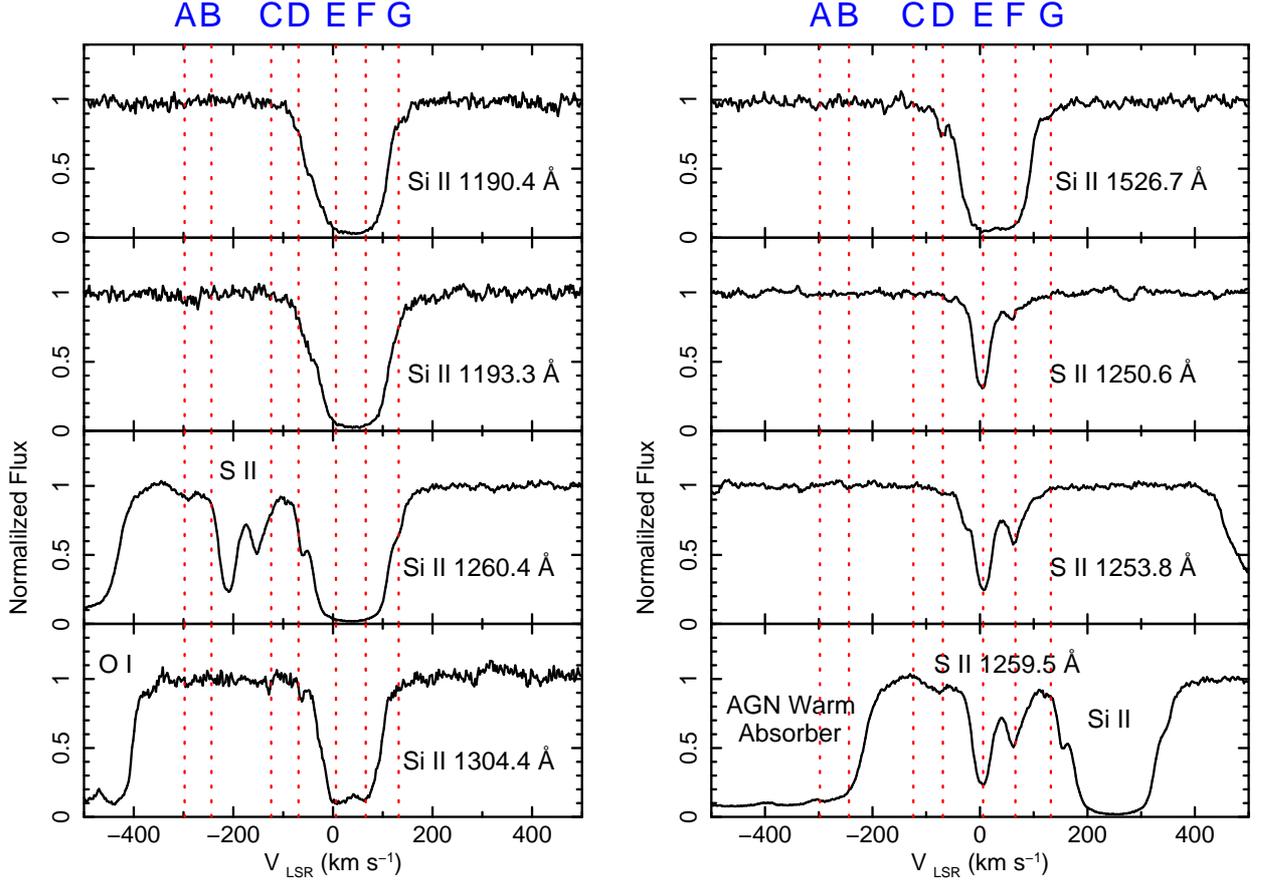}}
      \caption{HST/COS absorption lines from the cold phase (continued). Units are same as in Fig.~\ref{fig:cos_lines1}.}
          \label{fig:cos_lines2}
  \end{center}
\end{figure*}

\begin{figure*}
  \begin{center}
      \subfigure{ 
      \includegraphics[width=12cm, angle=-90]{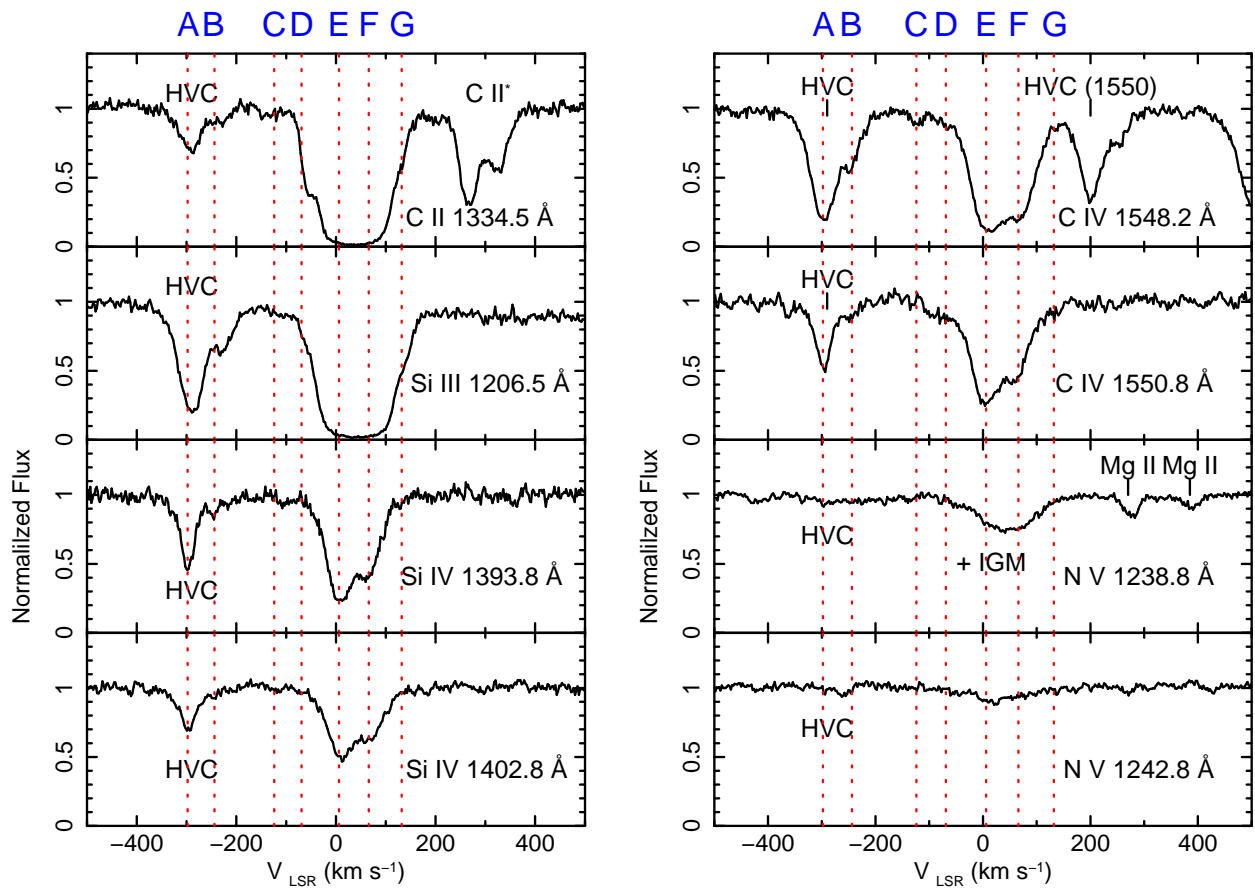}}
      \caption{HST/COS absorption lines from the warm phase {(ionization states III-V)}. Units are same as in Fig.~\ref{fig:cos_lines1}.}
          \label{fig:cos_lines3}
   \end{center}
\end{figure*}

Most features in the COS spectrum are foreground absorption lines due to the ISM of the Galaxy. We detected absorption lines from seven discrete ISM components (see Table~7, paper\,VI). We also found three absorption lines due to the diffuse intergalactic medium in the LOS towards the AGN.
In Figs. \ref{fig:cos_lines1}, \ref{fig:cos_lines2} and \ref{fig:cos_lines3} we plot the prominent interstellar lines present in the COS spectrum. The lines are sorted according to their ionization state and ionization potential. All lines are plotted in velocity space and in units of normalized flux. The zero-velocity of the scale is given by the laboratory wavelength of each transition.
% The velocity scale is defined by the well known Doppler formula
% \begin{equation}
%  v = \frac{\lambda - \lambda_0}{\lambda_0} \;c.
% \end{equation}
For consistency and easy comparison with previous work we then convert velocities to the Local Standard of Rest (LSR). The conversion from the laboratory to the LSR scale is given by v$_{\rm LSR}=\,$v$_{\rm Lab}+11.16\,$km\,s$^{-1}$ \citep{Blades1983}.

In the COS spectrum the ISM features are resonance lines and encompass up to 5 ionization states, e.g. from \ion{N}{i} to \ion{N}{v}. The strongest lines are the \ion{N}{i} triplet around 1200 {\AA}, the \ion{O}{i} line at 1302 {\AA}, and the \ion{Fe}{ii} and \ion{Al}{ii} lines at 1608.4 and 1670.8~{\AA}, respectively. Multiple strong transitions from both \ion{Si}{ii} and \ion{S}{ii} are shown, as well as \ion{C}{iv} and \ion{Si}{iv} doublets. The \ion{C}{ii} and \ion{Si}{iii} lines are heavily saturated. We confirm the presence of several discrete absorbers, seven in the case of \ion{C}{iv} as found in paper\,VI, with v$_{\rm LSR}$ of about -295, -240, -125, -65, +5, +65 and +90~to~+130 km\,s$^{-1}$. We label these absorbers with alphabetical letters ranging from A to G and sort them according to their v$_{\rm LSR}$ (Table~\ref{table:fit}). Moreover, we adopt the standard nomenclature for the clouds: the absorbers that produce components A, B and C are HVCs as their velocity modulus is greater than 100\,km\,s$^{-1}$. The absorbers responsible for components D, F, and G are intermediate velocity clouds (IVC). Component E is a low-velocity cloud (LVC).

\subsection{The FUSE spectrum}
\label{subsec:fuse_spectrum}

FUSE observed Mrk 509 in 1999 for about 52\,ks \citep{Kriss2000} and in 2000 for 62.1\,ks (paper\,VI). We use only the last observation because of its higher S/N ratio. We use the FUSE spectrum mostly for the analysis of the absorption lines due to the interstellar \ion{O}{vi} (see Fig. \ref{fig:warm_gas}).

\subsection{The RGS spectrum}
\label{subsec:rgs_spectrum}

The \textit{XMM-Newton} RGS spectrum of Mrk 509 is complex because most of the observed features are intrinsic to the AGN and some of them blend with foreground absorption lines as shown in paper\,II. Absorption and emission lines intrinsic to the AGN can be distinguished from the ISM lines as the former are significantly red-shifted. For a complete analysis of the AGN intrinsic absorption we refer to \citet{Detmers11}, hereafter paper\,III. The strongest ISM features are the \ion{O}{i} and \ion{N}{i} absorption lines at 23.5\,{\AA} and 31.3\,{\AA}, respectively. Clear evidence of highly ionized gas is provided by the \ion{O}{vii-viii} lines at 21.6 and 19.0\,{\AA}. We plot the individual RGS absorption edges of neutral O, Fe, and N together with the most prominent high-ionization lines in Fig.~\ref{fig:full_model}.

\section{Spectral modeling of the ISM}
\label{sec:analysis}

Our analysis focuses on the absorption lines of the ISM along the LOS towards Mrk 509. These lines are narrower than the features intrinsic to the AGN and always narrower than 1\,{\AA}. The only exception is the \ion{H}{i} Ly\,$\alpha$ absorption line at 1215.67\,{\AA}, which has a FWHM of $20$\,{\AA} (see Fig.~\ref{fig:cold_gas}). In order to reproduce both the AGN continuum and narrow and broad emission lines, we use the continuum emission model of paper\,VI. The spectral ranges that are strongly affected by intrinsic absorption were excluded as the AGN analysis is carried in other paper (Kriss et al. 2011). For the RGS spectral modeling we used the continuum plus line emission model (Model 2) given in paper\,III for the warm absorber.
First we analyze the ISM in the UV and in the X-rays separately due to the different resolution of the instruments. The different velocity components are resolved in the UV spectra (see Fig.~\ref{fig:cos_lines1}, \ref{fig:cos_lines2} and \ref{fig:cos_lines3}), but this it is not yet possible in the X-ray band.

\subsection{UV spectral fits}
\label{sec:cos_fits}

We have performed the UV analysis in two steps. First, we use an empirical model to estimate the velocity shift and dispersion of each absorption line and the ionic column density. This model provides information on the significance and kinematics of all the components. It further gives hints of their ionization state and location within the Galaxy. In the second method we use physical models to provide a more realistic description of the several foreground absorbers. These models provide important information on the ionization state, abundances and the sources responsible for heating the ISM. A final simultaneous fit to the UV and X-ray spectra was performed to obtain robust results by covering the entire range of ionization states.

\subsubsection{An empirical {\sl slab} model for the COS--UV spectra}
\label{sec:slab}

The first ISM model we use consists of seven {\sl slab} components, which are required in order to reproduce the 7 different kinematic ISM absorbers observed in the COS and FUSE spectra. Only \ion{C}{iv} clearly shows all seven components (see Fig.~\ref{fig:cos_lines3}). Excluding \ion{C}{ii}, the low-ionization absorption lines require no more than four components. Thus low-ionization gas is not detected for the HVCs. The {\sl slab} model in SPEX calculates the transmission of a slab of material, where all ionic column densities can be chosen independently. This has the advantage that the spectrum can be fitted without any prior knowledge of the ionization balance. The {\sl slab} model jointly fits all the lines that are are produced by the same ion. After an acceptable spectral fit is obtained, we can compare the observed column densities with those predicted from photo-ionization models. Free parameters in the {\sl slab} model are the velocity dispersion $\sigma_{\rm v}$, the Doppler velocity shift v and the ionic column density $N_{\rm X}$.

\begin{figure}[!hb]
\begin{center}
\includegraphics[width=13cm, angle=-90]{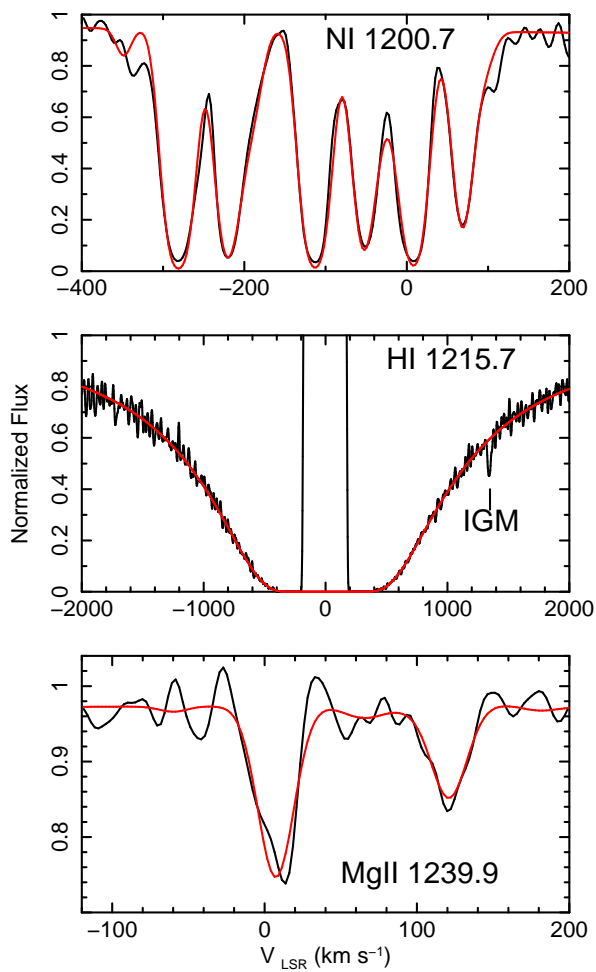}%}
\end{center}
\vspace{-.5cm}
\caption{Best fit to the HST/COS absorption lines of the cold gas with the {\sl slab} model (see also Table~\ref{table:fit}).}
\label{fig:cold_gas}
\end{figure}

For some UV saturated lines there is a degeneracy between the velocity dispersion and column density, hence the determined column density is uncertain. For several ions we only detect saturated lines in the UV spectrum, thus we have performed simultaneous fits for lines with similar ionization potential, like \ion{H}{i}, \ion{N}{i} and \ion{Mg}{ii} (see Fig.~\ref{fig:cold_gas}), tying their v and $\sigma_{\rm v}$. All lines of the same ion were simultaneously fitted like the four \ion{C}{i} lines (see Fig.~\ref{fig:cos_lines1}), the five \ion{Si}{ii} and three \ion{S}{ii} lines (see Fig.~\ref{fig:cos_lines2}), as well as the \ion{Si}{iv}, \ion{C}{iv} and \ion{N}{v} doublets (see Fig.~\ref{fig:cos_lines3}). In order to simplify our model and to shorten the CPU time we force the component A, B and C (HVCs) to have the same velocity dispersion, and similarly for components E, F and G (IVCs, see Table~\ref{table:fit}). A preliminary fit to the strong, resolved lines of \ion{N}{i}, \ion{Fe}{ii}, \ion{S}{ii}, \ion{C}{iv} and \ion{S}{iv} did give consistent values for the velocity dispersion $\sigma_{\rm v}$ within these groups. Only component D shows a rather smaller $\sigma_{\rm v}$, thus we treat it separately. In the cases where component D was too weak or blended, we froze $\sigma_{\rm v}$ to the value obtained for \ion{O}{i}, the strongest line in component D. The \ion{N}{v} 1238.8\,{\AA} line is affected by IGM absorption (paper\,VI) and the \ion{O}{vi} HVC lines are affected by the presence of molecular H$_2$ \citep{Wakker2006}. To reproduce the IGM line we added a slab of \ion{H}{i} with ${\rm v}=5700{\rm\,km\,s}^{-1}$, $\sigma_{\rm v}=40{\rm\,km\,s}^{-1}$ and very small column density $\log N_{\rm \ion{H}{i}} = 17.2\,{\rm m}^{-2}$ in agreement with paper\,VI. Two Gaussians were added to model the H$_2$ absorption at 1031.2\,{\AA}. These Gaussians have velocity shifts of +3 and +60\,${\rm\,km\,s}^{-1}$, $\sigma_{\rm v}=8{\rm\,km\,s}^{-1}$, and equivalent widths of 0.06 and 0.03\,{\AA}, respectively in agreement with  \citet{Wakker2006}. Thus, both the \ion{N}{v} and \ion{O}{vi} column densities, as given in Table~\ref{table:fit}, are corrected by IGM and H$_2$ contamination. The \ion{C}{ii} absorption lines appear to be different from the other low-ionization ions like \ion{S}{ii} and \ion{Si}{ii}. {Among the low ions,} only \ion{C}{ii} clearly shows high-velocity absorption. Moreover, a separate fit of the \ion{C}{ii} and \ion{C}{i} absorption lines provides a column density ratio \ion{C}{ii}/\ion{C}{i} $\sim100$ and a velocity dispersion ratio $\sigma_{\ion{C}{ii}}/\sigma_{\ion{C}{i}} > 3$, which are inconsistent with the other column density ratios of low ions like \ion{O}{i-ii} and \ion{N}{i-ii}. This suggests that the bulk of the neutral carbon in the cold phase is depleted into dust grains and that most of the \ion{C}{ii} is provided by the warm phase as confirmed by our physical models (see Sect.~\ref{sec:complete_model}.) For these reasons we prefer to fit the \ion{C}{ii} absorption lines together with the intermediate ions. We empirically model the \ion{H}{i} absorption assuming that it is all in the cold phase. Our physical model confirms that the the neutral hydrogen present in the cold phase is indeed two orders of magnitude larger than that in the warm and hot phases.

In Figs.~\ref{fig:cold_gas} and \ref{fig:warm_gas} we show the data and best fit models for the most prominent lines of the cold, warm and hot ISM phases. The results of the best fit with the seven slab components are reported in Table~\ref{table:fit}. They are sorted according to the velocity shift (columns) and ionization potential (rows). We also group the ions which have been fitted together: \ion{H}{i}, \ion{N}{i} and \ion{Mg}{ii}; \ion{Si}{ii} and \ion{Al}{ii}; \ion{Fe}{ii} and \ion{Ni}{ii}; \ion{C}{ii-iv}, \ion{Si}{iii-iv} and \ion{N}{v}. We generally detect low-ionization lines only for the LVC and IVCs (components E and D, F and G), thus four slab components are sufficient to model these low-intermediate velocity absorbers. However, we need seven slab components for modeling the mildly ionized gas providing the bulk of \ion{C}{ii} and all two or more times ionized ions. We note that the $\sigma_{\rm v}$ on average increases with the ionization state. A comparison of the column densities of the three HVCs (component A~to~C) shows that the fastest component A is the least ionized.

In order to simultaneously fit the absorption lines of the mildly ionized gas, we applied a few wavelength shifts to certain ions to match both the IVC and HVC features. These shifts are likely due to residual errors in the COS and FUSE wavelength calibration. We applied no shift for both \ion{C}{iv} and \ion{Si}{iv} as their profiles perfectly match (see Fig.~\ref{fig:warm_gas}). The \ion{Si}{iii} lines have been shifted by $-8{\rm\,km\,s}^{-1}$, the \ion{C}{ii} and \ion{N}{v} lines by $-10{\rm\,km\,s}^{-1}$. We also shifted the FUSE \ion{O}{vi} lines by $-14.5{\rm\,km\,s}^{-1}$ to obtain a match with the COS \ion{C}{iv} HVC lines. Because the \ion{O}{vi} features are blended and affected by H$_2$ absorption, in the spectral fits we prefer to fix all the \ion{O}{vi} LOS velocities to those of the \ion{C}{iv} and \ion{S}{ii} lines, which are well constrained.

\begin{table*}
\renewcommand{\arraystretch}{1.3}
\caption{Spectral fits to the COS and FUSE spectra using {the empirical model with} 7 slab components. Dashes are non detections.}
% \vspace{-0.75cm}
\begin{center}
% use packages: array
 \small\addtolength{\tabcolsep}{+2pt}
\scalebox{1.1}{%
\begin{tabular}{l|lll|l|lll}
\hline
           &   A$\,^{(a)}$&   B$\,^{(a)}$&   C$\,^{(a)}$&   D$\,^{(a)}$&     E$\,^{(a)}$&       F$\,^{(a)}$&   G$\,^{(a)}$\\
\hline
  $\sigma\,^{(b)}$    & ---     & ---    &  ---    & $\equiv 2.0\,^{(e)}$      &                 & $4.5\pm0.4$     &                \\
%     $v$       & ---     & ---    &  ---    & $-60\pm10$         & $-12\pm1$       & $51\pm1$        & $91\pm6$       \\
    $v\,^{(b)}$       & ---     & ---    &  ---    & $-50\pm10$         & $-1\pm1$       & $62\pm1$        & $100\pm10$       \\
 \ion{C}{i}$\,^{(c)}$   & ---     & ---    &  ---    & $15.7\pm0.3$      & $17.6\pm0.1$    & $16.8\pm0.1$    & $16.2\pm0.1$   \\
\hline
  $\sigma\,^{(b)}$    & $<10\,^{(d)}$    &  ---    &             ---   & $2.0\pm0.4$     &                 & $9\pm1$   &              \\
%     $v$       & $-294\pm7$       &  ---    & $-147\pm5$        & $-75.1\pm0.5$   & $-7\pm3$        & $44\pm5$      & $75\pm10$    \\
    $v\,^{(b)}$       & $-290\pm5$       &  ---    & ---               & $-64\pm1$       & $+7\pm3$        & $71\pm5$      & $131\pm10$    \\
 \ion{O}{i}$\,^{(c)}$   & $>16.4\,^{(d)}$  &  ---    & ---               & $19.2\pm0.4$    & $21.1\pm0.2$    & $20.7\pm0.1$  & $17.1\pm0.1$ \\
\hline
  $\sigma\,^{(b)}$    & ---     &  ---   &   ---   & $\equiv 2.0\,^{(e)}$      &                 & $7.5\pm0.5$     &                \\
%     $v$       & ---     &  ---   &   ---   & $-57\pm2$         & $-2.5\pm5.0$    & $57.5\pm0.2$    & $84.2\pm0.6$   \\
    $v\,^{(b)}$       & ---     &  ---   &   ---   & $-64\pm5$         & $+8\pm2$        & $70\pm1$        & $100\pm5$   \\
 \ion{H}{i}$\,^{(c)}$   & ---     &  ---   &   ---   & ---               & $24.53\pm0.02$  & $23.2\pm0.1$    & ---            \\
 \ion{N}{i}$\,^{(c)}$   & ---     &  ---   &   ---   & $16.4\pm0.4$      & $19.5\pm0.2$    & $18.6\pm0.1$    & $17.4\pm0.1$ \\
 \ion{Mg}{ii}$\,^{(c)}$ & ---     &  ---   &   ---   & $18.1\pm0.1$      & $19.8\pm0.1$    & $18.6\pm0.3$    & $18.2\pm0.1$   \\
\hline
  $\sigma\,^{(b)}$    & $<4\,^{(d)}$     &   ---   &   ---     & $\equiv 2.0\,^{(e)}$     &                 & $13\pm1$         &               \\
%     $v$       & $-310\pm2$       &   ---   &   ---     & $\equiv -82.6$   & $\equiv -6.3$   & $\equiv 50.5$    & $\equiv 78$   \\
    $v\,^{(b)}$       & $-300\pm10$      &   ---   &   ---     & $-65\pm5$         & $5\pm5$         & $65\pm5$         & $131\pm10$   \\
 \ion{Si}{ii}$\,^{(c)}$ & $16.7\pm0.2$     &   ---   &   ---     & $17.9\pm0.3$     & $19.5\pm0.4$    & $19.4\pm0.2$     & $16.8\pm0.3$  \\
 \ion{Al}{ii}$\,^{(c)}$ & $16\pm1$         &   ---   &   ---     & $17.3\pm0.1$     & $18.2\pm0.1$    & $17.5\pm0.2$     & $15.5\pm0.5$  \\
\hline
  $\sigma\,^{(b)}$    & $<7.6\,^{(d)}$   &   ---    &  ---     & $1.1\pm0.4$      &                 & $7.3\pm0.3$      &                \\
%     $v$       & $-314\pm5$       &   ---    &  ---     & $-83.3\pm0.3$    & $-8.4\pm0.2$    & $46\pm1$         & $77\pm1$       \\
    $v\,^{(b)}$       & $-300\pm5$       &   ---    &  ---     & $-72.1\pm0.3$    & $2.8\pm0.2$     & $58\pm1$         & $89\pm1$       \\
 \ion{Fe}{ii}$\,^{(c)}$ & $>16.7\,^{(d)}$  &   ---    &  ---     & $17.5\pm0.1$     & $19.3\pm0.1$    & $18.47\pm0.05$   & $17.56\pm0.03$ \\
 \ion{Ni}{ii}$\,^{(c)}$ & ---              &   ---    &  ---     & $16.7\pm0.2$     & $17.53\pm0.02$  & $17.30\pm0.03$   & $16.93\pm0.07$ \\
\hline
  $\sigma\,^{(b)}$    & ---     &  ---     & ---      & $3.6\pm0.6$      &                 & $6.4\pm0.7$      &                \\
%     $v$       & ---     &  ---     & ---      & $-82.6\pm0.3$    & $-6.3\pm0.1$    & $50.5\pm0.5$     & $78\pm2$       \\
    $v\,^{(b)}$       & ---     &  ---     & ---      & $-71.4\pm0.3$    & $4.9\pm0.1$    & $61.7\pm0.5$     & $89\pm2$       \\
 \ion{S}{ii}$\,^{(c)}$  & ---     &  ---     & ---      & $17.7\pm0.1$     & $20.6\pm0.1$    & $18.78\pm0.02$   & $17.8\pm0.1$   \\
\hline
  $\sigma\,^{(b)}$    &                & $12.0\pm0.2$   &                 & $2.6\pm0.6$   &                & $18\pm1$       &                \\
%     $v$       & $-308.6\pm0.2$ & $-253\pm1$     & $-128\pm6$      & $-80.8\pm0.4$ & $\equiv-6.3$   & $\equiv 50.5$  & $99\pm3$       \\
    $v\,^{(b)}$       & $-297.6\pm0.2$ & $-244\pm1$     & $-124\pm3$      & $-69\pm1$     & $6\pm1$        & $66\pm2$       & $132\pm5$       \\
 \ion{C}{ii}$\,^{(c)}$  & $17.62\pm0.03$ & $17.0\pm0.1$   & $<16.4\,^{(d)}$ & $19.2\pm0.5$  & $20.00\pm0.05$ & $19.10\pm0.05$ & $17.4\pm0.1$   \\
 \ion{Si}{iii}$\,^{(c)}$& $17.53\pm0.02$ & $16.59\pm0.03$ & $15.8\pm0.2$    & $16.0\pm0.1$  & $19.01\pm0.05$ & $18.40\pm0.06$ & $16.5\pm0.1$ \\
 \ion{Si}{iv}$\,^{(c)}$ & $17.34\pm0.01$ & $16.35\pm0.07$ & $15.6\pm0.3$    & $15.9\pm0.2$  & $17.70\pm0.01$ & $17.43\pm0.01$ & $15.9\pm0.1$   \\
 \ion{C}{iv}$\,^{(c)}$  & $18.26\pm0.01$ & $17.52\pm0.02$ & $16.7\pm0.2$    & $16.6\pm0.2$  & $18.34\pm0.01$ & $18.14\pm0.01$ & $16.9\pm0.1$   \\
 \ion{N}{v}$\,^{(c)}$   & $16.7\pm0.1$   & $16.5\pm0.2$   & $<16.2\,^{(d)}$ & $16.5\pm0.2$  & $17.30\pm0.07$ & $17.2\pm0.1$   & $<16.7\,^{(d)}$\\
\hline
  $\sigma\,^{(b)}$    &                & $14\pm2$       &                 & $6\pm4$       &                & $40\pm6$       &                \\
%     $v$       & $\equiv-308.6$ & $\equiv-253$   & $\equiv-128$    & $\equiv-80.8$ & $\equiv-6.3$   & $\equiv 50.5$  & $\equiv99$     \\
    $v\,^{(b)}$       & $\equiv-297.6\,^{(e)}$ & $\equiv-244\,^{(e)}$   & $\equiv-124\,^{(e)}$    & $\equiv-69\,^{(e)}$   & $\equiv 6\,^{(e)}$     & $\equiv 66\,^{(e)}$    & $\equiv132\,^{(e)}$     \\
 \ion{O}{vi}$^{(c)}$  & $17.6\pm0.1$   & $17.74\pm0.05$ & $17.5\pm0.1$    & $16.2\pm0.2$  & $18.25\pm0.03$ & $18.34\pm0.08$ & $17.3\pm0.1$   \\
\hline
\end{tabular}}
\label{table:fit}
\end{center}
% \vspace{-0.75cm}
$^{(a)}$ The different {\sl slab} components. \\
$^{(b)}$ Both the velocity dispersion $\sigma$ and the LSR $v$ are in units of km$\,$s$^{-1}$. \\
% $^{(c)}$ The $\sigma$ are taken the same in each group of components (A-B-C and E-F-G). \\
$^{(c)}$ The ionic column densities $N_{\rm X}$ are in log (m$^{-2}$) units. \\
$^{(d)}$ $2\,\sigma$ lower / upper limits, respectively. \\
$^{(e)}$ Fixed parameters.
% Shifts applied with respect to \ion{C}{iv}: -8\,km$\,$s$^{-1}$ to \ion{Si}{iii} lines, -10\,km$\,$s$^{-1}$ to \ion{N}{v} and \ion{C}{ii} lines, and -14.5\,km$\,$s$^{-1}$ to \ion{O}{vi} lines.
\end{table*}

A physical analysis of the interstellar absorbers requires more realistic models. In the next subsection we test two different SPEX models to reproduce both the neutral and ionized gas phases: the {\sl hot} and the {\sl xabs} models, which reproduce collisional and photo-ionization respectively.
% \vspace{-.5cm}

% \begin{figure*}
% \begin{center}
% \includegraphics[width=13cm, angle=-90]{lines_c2_upto_o6.ps}%}
% \end{center}
% \vspace{-.5cm}
% \caption{Simultaneous fit to the HST/COS and FUSE warm and hot gas components with the {\sl slab} model (see also Table~\ref{table:fit}).}
% \label{fig:warm_gas}
% \end{figure*}

\subsubsection{A collisionally-ionized model for the cold gas}
\label{sec:Probe_cold_gas}

The {\sl hot} model in SPEX calculates the transmission of a collisionally-ionized equilibrium (CIE) plasma. For a given temperature and set of abundances, the model calculates the ionization balance and then determines all the ionic column densities by scaling to the prescribed total hydrogen column density. At low temperatures this model mimics the neutral interstellar gas \citep[see SPEX manual and][]{kaastra09}. Free parameters in the {\sl hot} model are the hydrogen column density $N_{\rm H}$, the temperature $T$, the velocity dispersion $\sigma_{\rm v}$ and shift v and the abundances.

Following the ISM analysis of \citet{Pinto2010}, we have first modeled the cold gas with a low-temperature collisionally-ionized gas. The interstellar cold gas is responsible for the low ionization absorption lines: \ion{H}{i}, \ion{C}{i}, \ion{N}{i}, \ion{O}{i}, \ion{Mg}{ii}, \ion{Al}{ii}, \ion{Si}{ii}, \ion{S}{ii}, \ion{Ni}{ii}, and \ion{Fe}{ii}. These lines arise from the LVC and IVCs (components D-G) and no significant low-ionization absorption is detected at high outflow velocities (see Table~\ref{table:fit}). Thus, we modeled the cold gas with just four CIE components, one for the LVC and three for the IVCs. As previously assumed for the {\sl slab} model, we take the same $\sigma_{\rm v}$ for the three reddest absorbers, E-F-G. We also adopted the same abundances for components D~to~G: the poor statistics and the line blending give degenerate fits that do not allow us to constrain the abundances of these four absorbers separately. For the strongest components E and F we performed a fit with decoupled abundances for S, Fe and N. The abundances of both components were consistent within the errors.

At first we assumed proto-Solar abundances. The fit was not satisfactory for any N$_{\rm H}$ ($\chi^2_{\nu}>10$) due to the over-predicted ionic column densities: \ion{C}{i} $\sim3\times10^{20}\,{\rm m}^{-2}$, \ion{Si}{i} $\sim4\times10^{18}\,{\rm m}^{-2}$ and \ion{Mg}{ii} $\sim2.4\times10^{20}\,{\rm m}^{-2}$. The column densities as measured for both \ion{C}{i} and \ion{Mg}{ii} using the fit with the empirical slab models are much lower (see Table~\ref{table:fit}) and the \ion{Si}{i} lines are not detected in the spectrum.

A fit with free abundances provides only a partial improvements and the main problems are the high \ion{Si}{ii}/\ion{Si}{i} and \ion{C}{ii}/\ion{C}{i} column density ratios. These ratios can be reproduced only through the combined effect of a high gas temperature and depletion onto dust. A satisfactory fit is reached when we assume that at least 80\% of Mg and 90\% of C and Si are depleted from the gaseous phase into dust grains.

The weakly ionized gas of the LVC and IVCs shows rather high temperatures of about 15\,000--20\,000\,K as expected by the significant \ion{S}{ii} column density and the absence of \ion{S}{i} (which cannot be attributed to depletion into dust). This suggests that an important fraction of hydrogen is ionized. Component E, the one at rest, provides 90\% of the total $N_{\rm \ion{H}{i}}$, while component F is the second strongest but accounts for only $\sim$\,5\%. The measured $\sigma_{\rm v}$ are consistent with those estimated through the {\sl slab} model. The \ion{H}{i} column densities measured by the two different models are consistent with each other, but their total \ion{H}{i} column is only about 75\% of the value measured at 21\,cm \citep{Murphy1996}, which might be due to the high saturation of the UV line or most likely to the difference in beam-size between the 21 cm radio observation and the pencil-beam UV observation. Because the \ion{O}{i} 1302.2\,{\AA} line is heavily saturated, it is difficult to measure the oxygen abundance. Thus we freeze it to the proto-Solar value. For a complete analysis of the abundances we refer to Sect.~\ref{sec:rgs_fits}, in which we fit the entire UV / X-ray dataset. We have also tested an alternative photo-ionization model for the cold gas and obtained similar results (see Sect.~\ref{sec:alternative_models}).

\subsubsection{A photo-ionized model for the warm ionized gas}
\label{sec:Probe_warm_gas}

To model the warm (mildly) ionized gas responsible for the \ion{C}{ii-iv} and \ion{Si}{iii-iv} absorption observed in all velocity components, we tried a collisional ionization model. However, it is not possible to obtain a satisfactory fit with seven collisionally-ionized gas components for any of the HVCs (A~to~C) and LVC~/~IVCs (D~to~G). This is because collisional ionization results in a narrow peak in ionization, while a wide range of ionization states is observed. An alternative solution might be multi-phase CIE gas, which we have tested without obtaining satisfactory results (see Sect.~\ref{sec:alternative_models}), or cooling gas out of ionization equilibrium \citep[see e.g.][]{Gnat2007}. Here we use a photo-ionization model for the mildly ionized gas, which provides a good solution.

For the photo-ionization modeling we use the {\sl xabs} model in SPEX. The {\sl xabs} model calculates the transmission of a slab of material, where all ionic column densities are linked through a photo-ionization model. The relevant parameter is the ionization parameter $\xi = L / n r^2$, with $L$ the source ionizing luminosity between $1-1000$\,Ryd, $n$ the density and $r$ the distance from the ionizing source. Free parameters in the {\sl xabs} model are the hydrogen column density $N_{\rm H}$, the ionization parameter $\xi$, the velocity dispersion $\sigma_{\rm v}$, the Doppler velocity shift v, and the abundances. We have created a model consisting of seven photo-ionized {\sl xabs} components, one for each velocity component detected. Similar to the {\sl slab} and {\sl hot} models, we have coupled the $\sigma_{\rm v}$ and abundances within the two LVC~/~IVC and HVC groups (see Sect.~\ref{sec:slab} and Fig.~\ref{fig:warm_gas}), i.e. components A-B-C and E-F-G.

\begin{figure}
      \centering
      \includegraphics[angle=90, width=9cm]{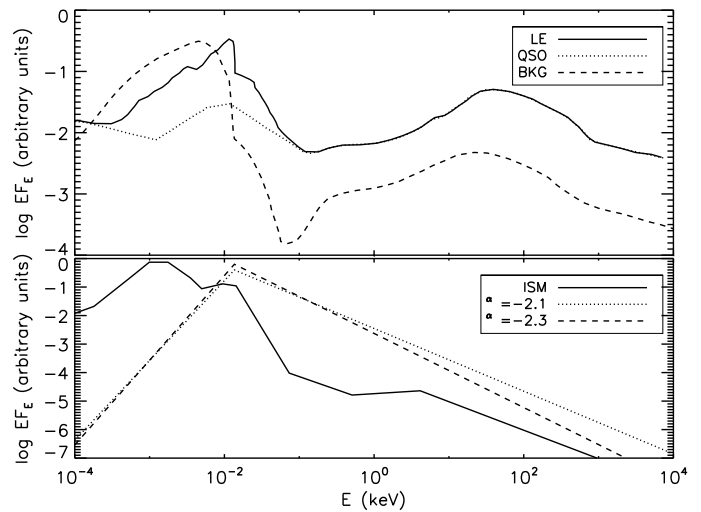}%}
      \caption{SEDs which have been used in the fit of the warm photo-ionized gas. Arbitrary units are used in order to compare the shape of the SEDs. LE SED refers to QSO + galaxies. See also Sect.~\ref{sec:Probe_warm_gas}.}
          \label{fig:SED}
    \end{figure}

The spectral energy distribution (SED) plays a crucial role in determining the ionization balance in the photo-ionized layers \citep[see e.g.][]{Chakravorty09}. Unfortunately, the SED which determines the ionization balance of the interstellar gas is not well known. For this reason, we have tested different SEDs on the seven ISM absorbers (see also Fig.~\ref{fig:SED}):
\vspace{-0.25cm}
\begin{enumerate}
 \item local emissivity (LE) of all the galaxies and QSOs, i.e. the integrated emission of galaxies and QSOs as seen in the local Universe at $z=0$;
 \item local emissivity (at $z=0$) of only QSOs;
 \item cosmic background radiation plus X-ray background as measured by HEAO1 and BeppoSAX;
 \item interstellar field SED (entirely due to starlight);
 \item powerlaw (PL) SEDs.
\end{enumerate}

\begin{figure*}
\begin{center}
\includegraphics[width=13cm, angle=-90]{lines_c2_upto_o6.ps}%}
\end{center}
\vspace{-.5cm}
\caption{Simultaneous fit to the HST/COS and FUSE warm and hot gas components with the {\sl slab} model (see also Table~\ref{table:fit}).}
\label{fig:warm_gas}
\end{figure*}

The first three SEDs represent a purely extragalactic ionizing source and have been taken from \citet{Haardt2001,Haardt2011}. The interstellar field SED refers to the one specified in the CLOUDY "Hazy" manual \citep{Ferland2005}. For the last case (SED 5) we have created a grid of powerlaw SEDs $(F_{\nu} \propto \nu^{\,\alpha})$ with slopes ranging from $-4.0$ to $-0.1$ with steps of 0.05 and with a low-energy cut-off below 1\,Ryd. We calculate the ionization balance through the SPEX {\sl xabsinput} tool: it receives as input the ionizing SED and the abundances of the gas, and determines the ionization balance using Cloudy version 08.00 (see the SPEX manual for more details).We have used the normalization for the physical SEDs as given in the literature, but we note that the absolute normalization of the SED does not matter while the SED shape does (see e.g. Sect.~6.2 in the SPEX manual). In order to have a sufficiently broad energy band, {\sl xabsinput} adds a minimum flux value at energies of $10^{-8}$ and $10^9$~Ryd, which does not affect the results. Because we do not know the abundances a priori, we have decided to fix them to the proto-Solar values of \citet{Lodders09}. We have adopted the same SED and abundances for the IVC group, and similarly for the HVC group. This choice is suggested by the fact that large deviations in ionization balance and metallicity are mostly expected by comparing LVC~/~IVCs lying in the Galactic disk with the halo HVCs (see below).

Among the four physically motivated SEDs, only the local emissivity (LE) SED including galaxies plus QSOs provides a satisfactory fit with proto-Solar abundances ($\chi_{\nu} \sim 1.3$, see Table~\ref{table:fit_xabs}) for all seven absorbers. The other physical SEDs do not provide good fits even when allowing for highly non-solar abundances. The powerlaw SEDs provide results similar to the best fitting LE SED. For absorption from both IVCs and HVCs the best fit is obtained by assuming a PL SED with slope $\alpha \sim -2$, and proto-Solar abundances. The main difference between the PL and LE SEDs are the derived $\xi$ values. The ionization parameters are systematically higher for a LE SED (see Table \ref{table:fit_xabs}). Component A is the least ionized of the three HVCs, component D is the least ionized among the four IVCs. Components E and F have the same ionization parameter. We find that HVCs are generally more ionized than the IVCs. {The conversion from the ionization parameter $\xi = L / n r^2$ (as usually defined in the X-rays) to the standard $U=n_{\gamma}/n_{\rm H}$ (commonly used at lower energies in HVC works) is not trivial and depends on the adopted SED. Examples of conversion factors are provided by \citet{Netzer2008}.}

As a next step we test for deviations from Solar abundances for both HVCs and IVCs. However, a complete set of accurate abundances can only be obtained by including the X-ray spectrum. Therefore we first discuss the components of the X-ray spectral models, before presenting a model that fits simultaneously FUSE, COS and RGS data and the abundances. We note that the error bars in Table~\ref{table:fit_xabs} are very small, especially for the column densities, due to the {quality} of the data. However, possible systematic errors, like those due to deviations from proto-Solar abundances, are not included (see e.g. the final complete model in Table~\ref{table:fit_physical}).

\begin{table*}
\renewcommand{\arraystretch}{1.3}
\caption{Best fit results for the COS spectrum using 7 photo-ionized absorbers with proto-Solar abundances.}
\begin{center}
% use packages: array
\small\addtolength{\tabcolsep}{-0pt}
\scalebox{1}{%
\begin{tabular}{ll|lll|l|lll}
\hline
SED type &         &      A$\,^{(a)}$          &      B$\,^{(a)}$          &      C$\,^{(a)}$         &     D$\,^{(a)}$           &   E$\,^{(a)}$           &     F$\,^{(a)}$         &     G$\,^{(a)}$    \\
\hline
%    $v$       & $-297.4\pm0.2$ & $-242\pm1$     & $-117\pm6$      & $-69.6\pm0.4$ & $\equiv 4.9$   & $\equiv 61.7$  & $110\pm5$       \\
% & v$_{\,\ion{C}{iv}}$ & $-308.8 f$      & $-254.6 f$      & $-124.1 f$     & $-80.54 f$      & $-6.3 f$      & $+50.2 f$     & $+93.8 f$\\
& v$_{\,\ion{C}{iv}}\,^{(b)}$ & $\equiv-297.6$  & $\equiv-244$    & $\equiv-124$   & $\equiv-69$     & $\equiv 6$    & $\equiv 66$   & $\equiv132$\\
\hline
\multirow{4}{*}{PL$^{(f)}$}
& $\sigma_v\,^{(b)}$            &                 & $11.1\pm0.1$    &                & $10.0\pm0.5$    &               & $19.0\pm0.1$  &          \\
& $N_{\,\rm H}\,^{(c)}$       & $0.44\pm0.01$   & $0.073\pm0.002$ & $0.012\pm0.002$& $0.030\pm0.002$ & $3.32\pm0.02$ & $2.05\pm0.02$ & $0.056\pm0.002$\\
& $\log\,\xi\,^{(d)}$         & $-0.56\pm0.01$  & $-0.51\pm0.01$  & $-0.51\pm0.03$ & $-3.00\pm0.03$  & $-1.70\pm0.01$& $-1.70\pm0.01$ & $-0.56\pm0.03$\\
& $\alpha_{\rm SED}$ $^{(e)}$ & $-2.3$          & $-2.3$          & $-2.3$         & $-2.1$          & $-2.1$        & $-2.1$        & $-2.1$     \\
\hline
\multirow{3}{*}{LE$^{(f)}$}
& $\sigma_v\,^{(b)}$            &                 & $9.6\pm0.1$     &                & $13.3\pm0.4$    &               & $18.8\pm0.1$  &          \\
& $N_{\,\rm H}\,^{(c)}$       & $0.54\pm0.01$   & $0.064\pm0.002$ & $0.009\pm0.001$& $0.033\pm0.001$ & $3.61\pm0.02$ & $2.39\pm0.03$ & $0.064\pm0.002$\\
& $\log\,\xi\,^{(d)}$         & $-0.07\pm0.01$  & $0.20\pm0.01$   & $0.17\pm0.08$  & $-2.5\pm0.1$    & $-1.10\pm0.01$& $-1.11\pm0.01$& $0.15\pm0.02$  \\
\hline
\end{tabular}}
\label{table:fit_xabs}
\end{center}
$^{(a)}$ The different photo-ionized {\sl xabs} components.\\
$^{(b)}$ Both the velocity dispersion $\sigma_v$ and $v_{\,\ion{C}{iv}}$ are in units of km$\,$s$^{-1}$. The $v_{\,\ion{C}{iv}}$ are fixed. \\
$^{(c)}$ The hydrogen column densities $N_{\rm H}$ are in $10^{23}$ m$^{-2}$ units.\\
$^{(d)}$ The ionization parameter $\xi = L / n_{\rm H} r^2$ is in units of $10^{-9} \, {\rm Wm}$.\\
$^{(e)}$ $\alpha_{\rm SED}$ is the best fitting slope for the PL SED (see also Sect.~\ref{sec:Probe_warm_gas}).\\
$^{(f)}$ PL refers to the fit assuming a power-law SED, LE refers to the Local Emissivity SED including galaxies and QSOs.
\end{table*} 

\subsection{X-ray spectral fits}
\label{sec:rgs_fits}

The RGS spectrum of Mrk 509 is shown in detail in paper\,III. Paper\,II lists the strongest interstellar lines in this spectrum. In our analysis of this spectrum we adopt the AGN continuum and emission line model given in paper\,III, as well as the outflow {\sl slab} model in order to subtract the absorption intrinsic to the AGN. The RGS X-ray spectrum is complementary to the COS-FUSE spectra, providing the column densities of both the weakly and mildly ionized O and Ne ions, which are important to constrain the warm gas. The hot gas mostly absorbs at X-ray wavelengths and can be thoroughly studied only in this energy domain. Moreover, some important absorption lines from neutral atoms are often saturated in the UV providing only lower limits to their column densities. In the X-rays the same ions usually provide non saturated lines, but with limited velocity resolution.

\subsubsection{An empirical {\sl slab} model for the RGS spectrum}
\label{sec:Toy_model}

We first determine the column densities independently of ionization balance as we also did for both COS and FUSE spectra. This is an important check of the column densities for those ions with saturated lines in the UV band, in particular \ion{O}{i} and \ion{N}{i}. It also gives column densities for ions absent in the UV spectra. Unfortunately, the seven interstellar components which are resolved in the UV spectra, are one blend in the RGS spectrum. This yields degeneracy when fitting the RGS spectrum and we propose the following solution. We create an empirical model with seven {\sl slab} components as in Sect. \ref{sec:slab}, but we freeze the column density ratios to those determined from the UV spectra for the seven velocity components for the cold, warm and hot phases. These column density ratios were determined from non-saturated UV lines. For the cold phase (\ion{O}{i-ii}) we use the \ion{Fe}{ii} and \ion{S}{ii} UV lines. For the warm phase (\ion{O}{iii-v}) we use \ion{C}{iv} and \ion{Si}{iv} lines. For the hot phase we use \ion{O}{vi}. It is thought that most \ion{O}{vi} arises from the conductive layer between the warm and the hot gas, but our physical models predict that in the LOS towards Mrk~509 at least half of the \ion{O}{vi} belongs to the hot gas and thus might be a possible indicator for it (see also Sect.~\ref{sec:alternative_models}). The column density ratios adopted are displayed in Table \ref{table:ratios}. The errors on these average ratios are estimated from the spread in the two column density ratios calculated; for instance, the error on the $N_{F}/N_E$ ratio for the cold gas is given by the difference in the respective ratios provided by the \ion{Fe}{ii} and \ion{S}{ii} columns. In the case of the hot phase we have adopted just the statistical errors on the \ion{O}{vi} column densities because it is the only highly ionized ion {(five times ionized or above)} present in the UV spectrum.
\begin{table*}
\renewcommand{\arraystretch}{1.3}
\caption{{Empirical} model for the RGS spectrum.}
\begin{center}
% use packages: array
 \small\addtolength{\tabcolsep}{+2pt}
\scalebox{1.1}{%
\begin{tabular}{l|lll|llll}
\hline
           & $N_{A}/N_E\,^{(a)}$  & $N_{B}/N_E\,^{(a)}$  & $N_{C}/N_E\,^{(a)}$  & $N_{D}/N_E\,^{(a)}$  & $N_{E}/N_E\,^{(a)}$  & $N_{F}/N_E\,^{(a)}$ & $N_{G}/N_E\,^{(a)}$   \\
\hline
    ${\rm v}$ (km$\,$s$^{-1}$)
              & $\equiv-298$  & $\equiv-244$  & $\equiv-124$  & $\equiv-69$     & $\equiv6$  & $\equiv66$    &   $\equiv132$       \\
\hline
cold phase    & 0             & 0             & 0             & $0.009\pm0.007$ & $\equiv1$  & $0.15\pm0.01$ & $0.02\pm0.01$ \\
warm phase    & $0.6\pm0.2$   & $0.10\pm0.05$ & $0.02\pm0.01$ & $0.017\pm0.001$ & $\equiv1$  & $0.58\pm0.05$ & $0.03\pm0.01$ \\
hot phase     & $0.22\pm0.07$ & $0.31\pm0.06$ & $0.18\pm0.06$ & $0.009\pm0.005$ & $\equiv1$  & $1.2\pm0.3$   & $0.11\pm0.04$ \\
\hline
\end{tabular}}
\label{table:ratios}
\end{center}
$^{(a)}$ Column density ratios for the cold, warm and hot phases used in the empirical model fit to the RGS spectrum. We normalize the ratios to component~E.
\end{table*}

The LSR velocities of the seven {\sl slab} components are fixed to the values measured from the UV \ion{C}{iv} lines. Moreover, we fix the velocity dispersion to the averages determined with UV {\sl slab} fits (see Table~\ref{table:fit}). In particular, for the low-ionization absorbers E, F and G we adopt a value of 9 km\,s$^{-1}$, which is the average $\sigma_{\rm V}$ of all the weakly ionized species. We did verify that a change of $\pm\,5$\,km\,s$^{-1}$ in $\sigma_{\rm V}$ (the average scatter in $\sigma_{\rm V}$) does not affect the column density estimates in the RGS fits.

All ionic column densities that are not constrained by the X-ray data, but have predicted X-ray continuum absorption, are fixed to the values calculated from the COS {\sl slab} fit. Examples are \ion{C}{i-iv} and \ion{Si}{ii-iv}. We also fix the \ion{H}{i} column densities because \ion{H}{i} only produces continuum absorption in the X-ray spectrum and there is no way to disentangle the different kinematics components. We split the total $N_{\rm \ion{H}{i}} = 4.44\times10^{24}$\,m$^{-2}$ \citep{Murphy1996} into four components representing D--G obtained above for the cold phase. The results of this model are listed in Table~\ref{table:empirical} and discussed in Sect.~\ref{sec:discuss_column}. We note that only the \ion{N}{i} column densities provided by the RGS fits are larger than those measured in the UV spectra, while the other column densities are consistent for the two different wavelength regions.

\begin{table*}
\renewcommand{\arraystretch}{1.3}
\caption{RGS spectral fits: empirical model with 7 slab absorbers. All the units are same as in Table~\ref{table:fit}.}
% \vspace{-0.75cm}
\begin{center}
% use packages: array
 \small\addtolength{\tabcolsep}{+2pt}
\scalebox{1.1}{%
\begin{tabular}{l|lll|l|lll}
\hline
      Par     &   A     &   B    &   C     &   D               &     E           &       F         &   G    \\
\hline
    $v$       & $\equiv-298$ & $\equiv-244$ & $\equiv-124$ & $\equiv-69$ & $\equiv 6$ & $\equiv 66$ & $\equiv132$ \\
\hline
  $\sigma$    & ---     & ---     & ---     & $\equiv 2.0$   &                & $\equiv 9$    &               \\
%  $f          $    & a     & b    & c     & d   &        1        & f    &      g       \\
 \ion{H}{i}   & ---     & ---     & ---     & $\equiv 22.2$  & $\equiv 24.6$  & $\equiv 23.6$  & $\equiv 22.6$ \\
 \ion{N}{i}   & ---     & ---     & ---     & $18.3$         & $20.27\pm0.06$ & $19.5$         & $18.6$ \\
 \ion{N}{ii}  & ---     & ---     & ---     & $16.7$         & $18.5\pm0.5$   & $17.8$         & $17.0$ \\
 \ion{O}{i}   & ---     & ---     & ---     & $19.2$         & $21.4\pm0.1$   & $20.5$         & $19.6$ \\
 \ion{O}{ii}  & ---     & ---     & ---     & $17.9$         & $19.8\pm0.3$   & $19.0$         & $18.2$ \\
 \ion{Ne}{i}  & ---     & ---     & ---     & $18.6$         & $20.7\pm0.1$   & $19.9$         & $19.0$ \\
 \ion{Ne}{ii} & ---     & ---     & ---     & $<18.4$ & $<20.4$ & $<19.6$ & $<18.7$ \\
 \ion{Mg}{i}  & ---     & ---     & ---     & $<18.7$ & $<20.7$ & $<19.9$ & $<19.0$ \\
 \ion{Mg}{ii} & ---     & ---     & ---     & $<18.7$ & $<20.7$ & $<19.9$ & $<19.0$ \\
 \ion{Fe}{i}  & ---     & ---     & ---     & $17.8$  & $19.7\pm0.3$   & $18.9$         & $18.1$         \\
 \ion{Fe}{ii} & ---     & ---     & ---     & $<17.8$ & $<19.8$ & $<19.0$ & $<18.1$ \\
\hline
  $\sigma$    &         & $\equiv 12$  &    & $\equiv 2.6$ &            & $\equiv 18$  &        \\
%  $f          $    & a     & b    & c     & d   &        1        & f    &      g       \\
 \ion{C}{v}   & $18.9$  & $18.1$  & $17.4$  & $17.3$       & $19.1\pm0.3$   & $18.9$   & $17.5$ \\
 \ion{O}{iii} & $19.2$  & $18.4$  & $17.6$  & $17.6$       & $19.5\pm0.2$   & $19.2$   & $17.8$ \\
 \ion{O}{iv}  & $19.0$  & $18.2$  & $17.5$  & $17.4$       & $19.3\pm0.3$   & $19.0$   & $17.6$ \\
 \ion{Ne}{iii}& $19.8$  & $18.9$  & $18.1$  & $18.1$       & $20.0\pm0.2$   & $19.7$   & $18.3$ \\
\hline
  $\sigma$    &        & $\equiv 14$ &         & $\equiv 6$ &              & $\equiv 40$ &      \\
%  $f          $    & a     & b    & c     & d   &        1        & f    &      g       \\
 \ion{C}{vi}  & $18.5$   & $18.7$   & $18.5$    & $17.1$   & $19.2\pm0.1$   & $19.3$   & $18.2$ \\
 \ion{N}{vi}  & $18.3$   & $18.5$   & $18.2$    & $16.9$   & $19.0\pm0.1$   & $19.1$   & $18.0$ \\
 \ion{N}{vii} & $18.2$   & $18.3$   & $18.1$    & $16.8$   & $18.9\pm0.3$   & $19.0$   & $17.9$ \\
 \ion{O}{vi}  & $18.2$   & $18.3$   & $18.1$    & $16.8$   & $18.9\pm0.3$   & $19.0$   & $17.9$ \\
 \ion{O}{vii} & $19.4$   & $19.6$   & $19.3$    & $17.9$   & $20.14\pm0.05$ & $20.2$   & $19.1$ \\
 \ion{O}{viii}& $19.0$   & $19.1$   & $18.9$    & $17.5$   & $19.7\pm0.1$   & $19.8$   & $18.6$ \\
 \ion{Ne}{ix} & $18.9$   & $19.0$   & $18.8$    & $17.4$   & $19.6\pm0.2$   & $19.7$   & $18.6$ \\
 \ion{Ne}{x}  & $18.5$   & $18.7$   & $18.4$    & $17.0$   & $19.2\pm0.5$   & $19.3$   & $18.2$ \\
 \ion{Fe}{xvii} & $16.6$ & $16.7$   & $16.5$    & $15.3$   & $17.2\pm0.2$   & $17.3$   & $16.3$ \\
\hline
\end{tabular}}
\label{table:empirical}
\end{center}
% \vspace{-0.75cm}
\end{table*}

\subsubsection{A self-consistent physical model of the ISM: simultaneous UV / X-ray spectral fits}
\label{sec:complete_model}

We construct a realistic ISM model consisting of a multi-phase structure by extending the UV ISM model. Essentially, we assume that the ISM consists of seven clouds or layers with different speeds but with a similar structure the LVC and IVCs (components D~to~G) have a cold phase of collisionally-ionized gas and molecules, a warm photo-ionized gas phase and a hot collisionally-ionized gas phase. The HVCs (components A~to~C) consist only of the warm and hot gas phases as suggested by the absence of high-velocity cold gas in the UV spectrum (see Table~\ref{table:fit_physical}). We simultaneously fit the RGS, COS and FUSE spectra in order to get the highest possible accuracy on line strength and broadening. UV saturated lines are ignored when lines from the same ion are not saturated in the X-ray spectrum. This is the case for the \ion{O}{i} and \ion{N}{i} UV lines at $+0\,$km\,s$^{-1}$ and $+65\,$km\,s$^{-1}$, which are saturated while their X-ray counterparts are not.

We take into account absorption by interstellar dust with the SPEX {\sl amol} component. The {\sl amol} model calculates the transmission of various molecules, for details see \citet{Pinto2010, Costantini2012} and the SPEX manual. The model currently takes into account the modified edge and line structure around the O and Si K-edge, and the Fe K / L-edges, using measured cross sections of various compounds taken from the X-ray literature.

The velocity dispersion $\sigma_{\rm v}$ is coupled within the component groups A-B-C and E-F-G as previously done (see Sect. \ref{sec:slab}). The $\sigma_{\rm v}$ of component D is still a free parameter as its lines are clearly narrower, especially for \ion{O}{i} (see Fig. \ref{fig:cos_lines1}). The LSR velocities are fixed to the values estimated by the \ion{C}{iv} {\sl slab} fits (see Table \ref{table:fit}). As in Sect. \ref{sec:Probe_cold_gas}, the abundances of each gas phase are coupled within component groups A-B-C (HVCs) and D-E-F-G (IVCs). For components C, D and G it is very difficult to measure accurate independent abundances because of their weak and blended profiles, while this is possible for components A, B, E, and F. However, we also do not want to increase the complexity of our model unnecessarily. Thus, we separately fitted the profiles of these latter four strongest components and found that the abundances of components A and B were in good agreement, although less constrained, and the same applies for components E and F. We also decided to keep proto-Solar abundances for all the HVCs because a fit with free abundances does not provide strong constraints besides Si/C$\gtrsim$1. Because in terms of $\chi^2_{\nu}$ the physical model provides results as good as those obtained with the empirical {\sl slab} model, we prefer to maintain our choice of abundances.

\subsubsection{Metal depletion}
\label{sec:Probe_abundances}

\begin{figure*}
    \centering
%       \subfigure[RGS - oxygen edge]{ 
      \includegraphics[angle=-90, width=18cm]{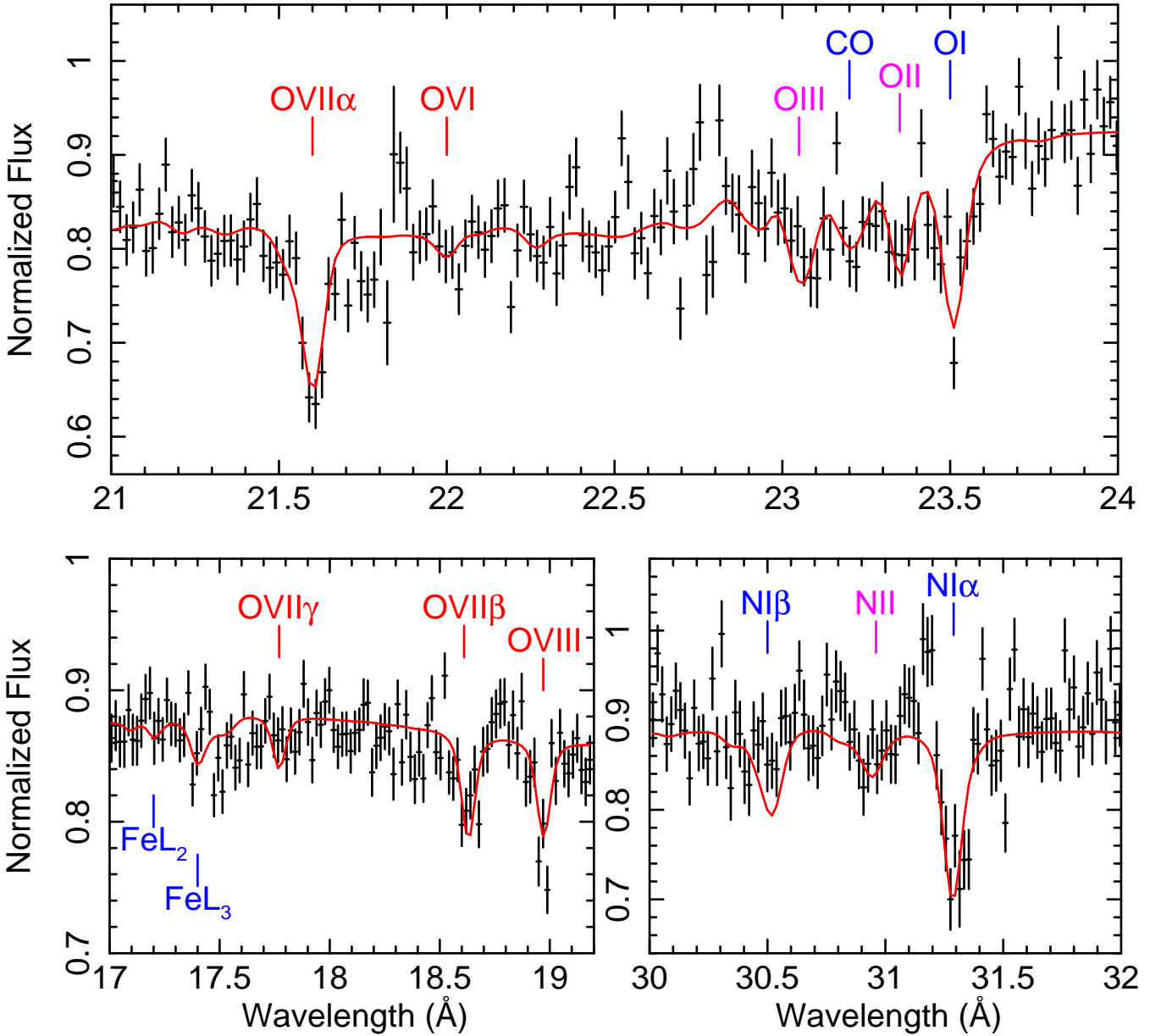}%}
\caption{Mrk 509 RGS data and best fit for the complete ISM model (see Sect.~\ref{sec:complete_model}). All the absorption and emission lines intrinsic to the AGN have been subtracted through the AGN {\sl slab} model of \citet{Detmers11}. The flux is normalized to the AGN continuum emission. Color code of the labels: red for the hot phase, purple for the warm phase and blue for the cold phase.}
    \label{fig:full_model}
    \end{figure*}

If the cold and warm phases share the same Galactic environment, but have different heating processes, their abundances might be still similar. However, the ratios of the column density estimates that we have reported in Table \ref{table:fit} and \ref{table:empirical} show strong discrepancies between neutral and ionized gas column densities. In particular the \ion{C}{i} column density is at least two orders of magnitude smaller than that for \ion{C}{ii}, while \ion{O}{i-ii} and \ion{N}{i-ii} do not show similar trends. It is unlikely that in the cold gas the carbon abundance is orders of magnitude smaller than Solar. The most reasonable explanation is instead that most of the neutral carbon is depleted in dust grains or molecules. The same applies for Fe, Si, Al. There are indeed no detections in the UV spectrum of any of their neutral transitions. Moreover, the Fe column densities are poorly constrained by the RGS {\sl slab} model (see Table~\ref{table:empirical}) due to the weakness of the Fe L-edge (Fig.~\ref{fig:full_model}). The column densities estimated for oxygen and nitrogen with the diagnostic model are instead close to those predicted for proto-Solar abundances, which suggests that their depletion factors should be much lower than those for C and Fe. We can measure the depletion factors by testing for the presence of CO, H$_2$O ice, silicates and other molecules through the {\sl amol} model adopted. The depletion of the cold gas phase into dust grains argues in favor of not coupling the abundances of the cold and warm gas for those elements which are expected to be involved like C, O, Mg, Al, Si, Fe, Ni. Instead we couple the abundances of N, Ne and S of the cold and warm gas because these elements are not expected to be highly depleted \citep{Wilms}.

\subsubsection{A collisionally-ionized model for the hot ionized gas}
\label{sec:Probe_hot_gas}

The main difference between the UV ISM model and the final ISM model is the addition of seven collisionally-ionized gas {\sl hot} components, one for each layer. They will reproduce the highly-ionized gas that we have previously probed with the diagnostic empirical model, see Table~\ref{table:empirical}. The only free parameters for each component are the hydrogen column density, the temperature and the velocity dispersion. The Doppler velocities are coupled according to the prescriptions given above. However, here we couple the velocity dispersion of components D--G as they fully blend in all the spectra. A satisfactory fit for the hot gas is reached by assuming proto-Solar abundances for all the seven {\sl hot} components. Indeed, a fit with free abundances does not provide any significant deviation from the abundances of \citet{Lodders09}. For this reason we report the results obtained by assuming proto-Solar abundances for the hot gas in Table~\ref{table:fit_physical}. We find that most of the hot gas is at rest and originates from the the slow components E and F. We obtain only upper limits to the column densities of the highly-ionized gas for the other components. We discuss these results in Sect.~\ref{sec:discuss_column}.

\begin{table*}

\renewcommand{\arraystretch}{1.3}

\caption{UV / X-ray simultaneous fits with the final complete ISM model.}

\begin{center}
% use packages: array
 \small\addtolength{\tabcolsep}{-1pt}
\scalebox{1}{%
\begin{tabular}{l|l|lll|llll}

\hline
 Phase &  Par  &   A            &   B            &   C             &   D           &     E          &       F        &   G            \\
&    v$\,^{(a)}$ & $\equiv-297.6$ & $\equiv-244$   & $\equiv-124$    & $\equiv-69$   & $\equiv 6$     & $\equiv 66$    & $\equiv132$    \\
\hline
\multirow{11}{*}{Cold gas}
&  $\sigma_v\,^{(a)}$    & ---            & ---            &  ---            & $<2$  &                & $9.6\pm0.1$    &                \\
&  $N_{\rm H}\,^{(b)}$   & ---            & ---            &  ---            & $0.9\pm0.1$   & $44\pm1$       & $8\pm1$        & $0.09\pm0.04$  \\
&  T (eV)                & ---            & ---            &  ---            & $<0.6$& $1.26\pm0.01$  & $1.24\pm0.02$  & $2.7\pm0.1$    \\
&  C/H$\,^{(c)}$         & ---            & ---            &  ---            &               \multicolumn{4}{c}{$<0.1$}                 \\
&  N/H$\,^{(c)}$         & ---            & ---            &  ---            &               \multicolumn{4}{c}{$0.8\pm0.1$}                    \\
&  O/H$\,^{(c)}$         & ---            & ---            &  ---            &               \multicolumn{4}{c}{$0.5\pm0.1$}                    \\
&  Ne/H$\,^{(c)}$        & ---            & ---            &  ---            &               \multicolumn{4}{c}{$1.3\pm0.1$}                    \\
&  Mg/H$\,^{(c)}$        & ---            & ---            &  ---            &               \multicolumn{4}{c}{$0.41\pm0.04$}                  \\
&  Al/H$\,^{(c)}$        & ---            & ---            &  ---            &               \multicolumn{4}{c}{$0.06\pm0.01$}                \\
&  Si/H$\,^{(c)}$        & ---            & ---            &  ---            &               \multicolumn{4}{c}{$0.14\pm0.04$}                  \\
&  S/H$\,^{(c)}$         & ---            & ---            &  ---            &               \multicolumn{4}{c}{$0.67\pm0.01$}                  \\
&  Fe/H$\,^{(c)}$        & ---            & ---            &  ---            &               \multicolumn{4}{c}{$0.07\pm0.01$}                  \\
&  Ni/H$\,^{(c)}$        & ---            & ---            &  ---            &               \multicolumn{4}{c}{$0.07\pm0.01$}                  \\
\hline
\multirow{9}{*}{Warm gas}
&  $\sigma_v\,^{(a)}$    &                & $9.8\pm0.1$    &                & $4.7\pm0.2$   &                & $18.5\pm0.1$   &               \\
&  $N_{\rm H}\,^{(b)}$   & $0.52\pm0.01$  & $0.065\pm0.002$&$0.010\pm0.002$ &$0.034\pm0.002$& $5.4\pm0.3$    & $2.9\pm0.1$    & $0.013\pm0.002$\\
&  $\log \xi\,^{(b)}$    & $-0.03\pm0.01$ & $0.17\pm0.01$  & $0.22\pm0.08$  & $-2.3\pm0.2$  & $-1.13\pm0.01$ & $-1.12\pm0.01$ & $0.7\pm0.2$  \\
&  C/H$\,^{(c)}$         & \multicolumn{3}{c|}{$\equiv 1$  }                &  \multicolumn{4}{c}{$\equiv 1$}                 \\
&  O/H$\,^{(c)}$         & \multicolumn{3}{c|}{$\equiv 1$  }                &  \multicolumn{4}{c}{$0.5\pm0.1$}                \\
&  Al/H$\,^{(c)}$        & \multicolumn{3}{c|}{$\equiv 1$  }                &  \multicolumn{4}{c}{$0.33\pm0.08$}              \\
&  Si/H$\,^{(c)}$        & \multicolumn{3}{c|}{$\equiv 1$  }                &  \multicolumn{4}{c}{$0.60\pm0.01$}              \\
&  Fe/H$\,^{(c)}$        & \multicolumn{3}{c|}{$\equiv 1$  }                &  \multicolumn{4}{c}{$0.90\pm0.06$}               \\
\hline
\multirow{3}{*}{Hot gas}
&  $\sigma_v\,^{(a)}$    &                & $<11$  &                 &  \multicolumn{4}{c}{$16\pm6$}                 \\
&  $N_{\rm H}\,^{(b)}$   & $<0.3$ & $<0.4$ & $<0.5$  & $1.7\pm0.5$   & $19\pm3$       & $4\pm1$        & $<1$ \\
&  T (eV)                & $40\pm10$      &$55\pm15$       & $51\pm14$       & $14\pm1$      & $160\pm10$     & $70\pm10$      & $55\pm15$    \\
\hline
\multirow{3}{*}{Molecules}
&  CO$\,^{(d)}$          & ---     & ---    &  ---    &  ---  & $7.5\pm1.5$        & ---    &  ---   \\
&  H$_2$O ice$\,^{(d)}$  & ---     & ---    &  ---    &  ---  & $<1.0$ & ---    &  ---   \\
&  MgSiO$_3\,^{(d)}$   & ---     & ---    &  ---    &  ---  & $<0.8$ & ---    &  ---   \\
&  Metallic Fe$\,^{(d)}$ & ---     & ---    &  ---    &  ---  & $<0.6$ & ---    &  ---   \\
&  Fe$_2$O$_3\,^{(d)}$ & ---     & ---    &  ---    &  ---  & $<0.3$ & ---    &  ---   \\
&  H$_2\,^{(d,\,e)}$          & ---     & ---    &  ---    &  ---  & $10-50$        & $4-6$    &  ---   \\
\hline

\end{tabular}}
\label{table:fit_physical}
\end{center}
$^{(a)}$ Velocity units are same as in Table \ref{table:fit}. Doppler velocities $v$ are fixed.\\
$^{(b)}$ The hydrogen column densities $N_{\rm H}$ are in units of $10^{23}$ m$^{-2}$. The ionization parameter $\xi$ is in units of $10^{-9} \, {\rm Wm}$.\\
$^{(c)}$ Abundances ratios are {in the linear proto-Solar abundance} units of \citet{Lodders09}.\\
$^{(d)}$ Molecular column densities are in units of $10^{20}$ m$^{-2}$.\\
$^{(e)}$ H$_2$ column densities are taken from \citet{Wakker2006}.
\end{table*}

\subsection{Alternative models}
\label{sec:alternative_models}

In this section we report several tests to check the validity and uniqueness of our model.

The first issue concerns the physical state of the cold weakly ionized gas. We have used a collisionally-ionized model in SPEX because at low temperatures this mimics well the cold interstellar gas (see Sect.~\ref{sec:Probe_cold_gas}). However, at these temperatures photo-ionization could also provide a significant contribution. Thus, we tried an alternative model in which four additional photo-ionized {\sl xabs} components (E~to~G) substitute for the cold collisionally-ionized ones. The fit slightly worsens and absolute abundances, i.e. relative to hydrogen, are not well constrained. They show systematic deviations of about 25--50\% from the values obtained with the collisionally-ionized model, while abundances relative to oxygen or other well constrained ions have smaller deviations. In summary, a photo-ionized plasma provides a worse fit to the cold phase, but in principle part of the cold component might be photo-ionized.

Excluding \ion{C}{ii}, ions like \ion{O}{i-ii} and \ion{O}{iii-iv} have been put in two different phases in the {\sl slab} model (see Sect.~\ref{sec:slab}). Although the physical model takes into account contributions to each ion from any phase, we have to justify the choice adopted for the empirical {\sl slab} model. At first we chose to fit the \ion{C}{ii} together with the \ion{C}{iv} and \ion{Si}{iii-iv} because it was the only `low' ion clearly showing high-velocity absorption. The physical model validates this choice as most of the carbon in the cold phase turns out to be locked into dust and the bulk of \ion{C}{ii} comes from the warm photo-ionized phase. For completeness we have tested a photo-ionization model for each velocity component in order to fit the entire set of neutral to doubly ionized ions (and ignoring all the other lines), assuming they are from a warm weakly-ionized phase and decoupled from the remaining more highly ionized ions. However, the fit is worse than before $(\Delta\chi^2_{\nu}\sim+1.0)$; the observed \ion{C}{ii}/\ion{C}{i}, \ion{Si}{ii}/\ion{Si}{iii} and \ion{S}{ii}/\ion{S}{iii} column density ratios are not well reproduced. Even if we add dust depletion for carbon and silicon we cannot obtain satisfactory results (at any N$_{\rm H}$).

We have also considered an alternative interpretation of the warm mildly-ionized gas that produces most of \ion{C}{iii-iv} and \ion{Si}{iii-iv}. Instead of seven photo-ionized absorbers (see Sect.~\ref{sec:Probe_warm_gas}) we used multiphase collisionally-ionized gas: each velocity component has been fitted with 2 or even 3 CIE components. This model provides the worst fit so far. It might be that the gas is out of equilibrium, but we cannot check this further with the current spectral models in SPEX. However, the goodness of fit of our standard model suggests that photo-ionization is a likely interpretation.

The last important issue concerns \ion{O}{vi}. We have so far considered this ion, together with \ion{O}{vii-viii}, as a tracer of the hot highly-ionized gas. However, as we have previously mentioned in Sect.~\ref{sec:Toy_model}, most of \ion{O}{vi} is thought to arise from a cooling colder phase with temperatures $1-5\times 10^5$\,K \citep[see e.g.][]{Richter2006}. Therefore, we have applied a collisionally-ionized model for this conductive phase that produces \ion{N}{v} and \ion{O}{vi} assuming a temperature of $2.1-4.4\times10^{-2}$\,keV ($2.5-5.1\times 10^5$\,K). However, this model is incompatible with another collisionally-ionized phase producing \ion{O}{vii-viii}. In order to explain the \ion{O}{viii}/\ion{O}{vii} and \ion{Ne}{x}/\ion{Ne}{ix} ratios this highest-ionization gas also provides at least half of the observed \ion{O}{vi}. In summary, either half of the \ion{O}{vi} is contained in the hot ionized gas, or a significant amount of \ion{O}{vii} belongs to the cooling intermediate phase. In both cases there is a link between the column densities of the \ion{O}{vi-vii} ions. For this reason we prefer our standard model and fit together all highly-ionized ions including and exceeding those five times ionized, like \ion{O}{vi}.

\section{Discussion}
\label{sec:discussion}

Our analysis shows that the ISM in the LOS of Mrk\,509 has a multi-phase structure (see Table \ref{table:fit} and \ref{table:fit_physical}) and complex dynamics. The ISM structure is similar for the different velocities components A--G, but the LVC and IVCs show a three-phase structure while the HVCs show only two ionized phases. We first discuss the general structure of the ISM that we have probed, its constituents and chemistry. Then, we will characterize the several absorbers, describe their heating processes and locate them in the Galactic environment.

\subsection{ISM multi-phase structure}
\label{sec:ISM_structure}

In the LOS towards Mrk\,509 the ISM is found in different physical and chemical forms. We found the gas at various temperatures with different heating processes. Our UV spectral modeling has revealed a large sample of ionization states (see Table \ref{table:fit}) which is further enlarged by including the X-ray detected absorption. The velocity dispersion is the same within clouds which belong to the same family, such as the IVCs. We have indeed obtained a very good fit by coupling the velocity dispersion of the LVC and IVC components E, F, and G (see Figs. \ref{fig:cold_gas} and \ref{fig:warm_gas}). Component D is an exception.

\subsubsection{The cold phase}
\label{sec:disc_cold_phase}

We have successfully modeled the interstellar gas with three main phases (see Table~\ref{table:fit_physical}). For the cold gas we have used a collisionally-ionized gas with low temperatures $kT \sim 0.5-2.7$\,eV {($6\,000-30\,000$\,K)}. It provides the bulk of the neutral and low-ionization lines, such as \ion{O}{i-ii}, \ion{N}{i-ii}, \ion{Fe}{i-ii} and \ion{S}{ii}. We detect the cold gas only for the LVC and IVCs, no neutral gas is found at high speed (see Table~\ref{table:fit}). More than 90\% of the neutral gas is provided by the LVC component~E. Component F (V$_{\rm LSR}$ = 65 km s$^{-1}$) is the second in order of column density and has an ionization state consistent with the LVC (for both the cold and warm gas), which suggests a similar environment for the two clouds.

In Sect.~\ref{sec:Probe_cold_gas} we have already shown that the \ion{C}{i} column densities are too low with respect to those of \ion{O}{i} and \ion{N}{i} for the \ion{C}{i}~/~\ion{C}{ii} ratio to be explained only by heating of the cold gas phase. We also argued that at least 90\% of C and Si from the cold phase is locked into molecules and dust grains. We confirm this with the complete physical model detailed in Table~\ref{table:fit_physical}. We have indeed obtained a 2$\sigma$ upper limit of 0.1 for both the carbon and silicon abundances with respect to the proto-Solar value of \citet{Lodders09}. This strongly suggests the presence of both carbonaceous and silicate compounds in the LOS. We have thus tested all the molecules available in the SPEX database and found the best match using a significant column of CO (note the feature at 23.2\,{\AA} in Fig.~\ref{fig:full_model}). We get only upper limits for metallic iron, pyroxene (MgSiO$_3$), and hematite (Fe$_2$O$_3$). We are only able to detect dust at rest (component E). Moreover, the broad profile of dust and its nearness to AGN intrinsic features makes the detection of weak lines due to dust at different velocities difficult in the RGS spectrum. It is possible to estimate the CO/H$_2$ ratio in the LOS of Mrk\,509 by comparing our CO column density estimate with the H$_2$ value measured by \citet{Wakker2006}, see also Table~\ref{table:fit_physical}. We obtain CO/H$_2 \sim 0.3\,(0.1-0.6)$, which is unusually high for the diffuse ISM. However, we cannot be highly confident in this value as the CO line falls in a wavelength range which is strongly affected by emission and absorption features intrinsic to the AGN (see paper\,III).

\subsubsection{The warm phase}

The warm phase consists of mildly-ionized gas in photo-ionization equilibrium. It spreads between several ionization states as it provides most \ion{C}{ii-iv} and \ion{N}{v}. In Sect.~\ref{sec:Probe_warm_gas} we tested several reasonable SEDs as photo-ionizing source for all the seven clouds. We obtained satisfactory fits only by taking into account the contribution from QSOs together with the extragalactic emission at $Z=0$ \citep{Haardt2011}. The fit with the SED containing only stellar light is not acceptable because this SED is not able to produce the high \ion{C}{iv} columns, which we have independently measured with the {\sl slab} model (see Table~\ref{table:fit}). The total local emissivity (galaxies plus quasars, LE) is the only SED able to photo-ionize the interstellar gas up to the level observed. We have also tested a diagnostic power-law (PL) SED and constrained ionization parameters $\xi$ which are systematically lower than those estimated with the physical LE SED (see Table~\ref{table:fit_xabs}). This might be due to the excess in softer part of the LE with respect to the PL SED caused by the additional extragalactic emission. The warm gas is the interstellar component which is best detected at high velocities (see Fig.~\ref{fig:cos_lines3}).

\subsubsection{The hot phase}

The hot phase is characterized by highly-ionized gas with temperatures of $50-160$~eV, i.e. $0.58-1.9\times10^6$\,K, (see Table~\ref{table:fit_physical}). It is responsible for the entire \ion{O}{vi-viii}, \ion{N}{vi-vii}, \ion{Ne}{ix} and \ion{C}{vi} absorption (Fig.~\ref{fig:full_model}). We detect hot gas in both HVCs, LVC and IVCs through the UV \ion{O}{vi} lines (see Fig.~\ref{fig:warm_gas}). Most of the hot gas is provided by the slow components E and F. The {hot phases of the} IVCs and HVCs show lower temperatures, closer to that of an \ion{O}{vi} interface between the warm and the hot gas, while the temperature of the LVC {hot phase} is in agreement with that of the typical hot coronal gas of the Galaxy \citep[see e.g.][]{Yao2005}. In X-rays components A-C merge with D-G and we can only measure upper limits for the column densities of the individual kinematic components (see Table~\ref{table:fit_physical}). We have measured large \ion{O}{vi} column densities in the FUSE spectrum (Table~\ref{table:fit}) for components A-C. The high column densities of ionized gas and the non-detection of neutral gas suggest that components A-C should be mostly or entirely ionized by both UV/X-ray background and collisions with the halo of our Galaxy or the Local Group.

In order to estimate the location of the hot gas of components E and F we perform a test on the 620\,ks MOS 1-2 data (paper\,I). We selected an annular region between 10-12' around Mrk\,509 in both the MOS detectors and obtained their spectra. We have fitted the 0.5--1.0\,keV MOS spectra with a power-law continuum and three Gaussians to describe the \ion{O}{vii-viii} and \ion{N}{vii} emission lines. The counts and fluxes are corrected for vignetting. The line fluxes in photons\,m$^{-2}$\,s$^{-1}$\,sr$^{-1}$ are: $f_{\ion{O}{vii}}=(4.50\pm0.35)\times10^4$, $f_{\ion{O}{viii}}=(1.39\pm0.15)\times10^4$, $f_{\ion{N}{vii}}=(1.11\pm0.12)\times10^4$. From the \ion{O}{viii}\,/\,\ion{O}{vii} line ratio we estimate that the hot gas has an average temperature of $0.186\pm0.006$\,keV, which is close to that of component E (see Table~\ref{table:fit_physical}).
For a source with a solid angle of 1\,sr, at a nominal distance of $10^{22}$\,m, with the observed flux given by the \ion{O}{viii} lines, at the measured temperature above, we obtain a CIE emission measure $Y=1.024\times10^{70}$\,m$^{-3}$.
The emission measure is also given by $Y=(n_{\rm e}/n_{\rm H})\,n_{\rm H}^2\,A \,dR$, where $A=10^{44}\,$m$^2$ is the surface area, $V=A\,dR$ the volume, $n_{\rm e}/n_{\rm H}=1.198$, and $dR$ is the depth of the gas layer.
We derive $n_{\rm H}^2\,dR = (8.5\pm0.9)\times 10^{25}$ m$^{-6}\,$m, which does not depend on the adopted distance.\\
\textbf{We consider three scenarios -}
In our first scenario, the gas emitting \ion{O}{vii} and \ion{O}{viii} is the same gas responsible for the high ionization absorption. This gas then provides the column density that we have measured for the absorption (Table~\ref{table:fit_physical}) and the emission measure $n_{\rm H}^2\,dR$ and because $N_{\rm H} = n_{\rm H}\,dR$ we can determine the thickness of the gas layer $dR=1.6\times10^{6}$\,pc and its density $n_{\rm H}=40$\,m$^{-3}$. This would suggest that the hot gas is related to the warm-hot intergalactic medium (WHIM). So, it would be physically decoupled from the other two (cold and warm) ISM phases. In Sect.~\ref{sec:IVCs_location} we show that components E--G can be assigned to the Galactic disk. The broad and unresolved profiles of the high ionization \ion{O}{vi-viii} lines are blended. As a test, we have fitted all the high-ionization lines with the velocities free to vary. In this fit only two IVCs are required (E and F) with $v_{\rm LSR}=$\,30 and 95\,km\,s$^{-1}$, $kT=$\,65 and 185\,eV, $N_{\rm H}=$\,9 and 6\,$\times10^{23}$\,m\,$^{-2}$, respectively. Most of the \ion{O}{vii-viii} is produced by the 185\,eV gas, whose column density provides a depth of 700\,kpc according to the equations mentioned above. In this case the hot gas should be embedded in the halo of the Local Group. {This means that the $1.6$\,Mpc value determined above is not plausible.}

In our second scenario, the emitting and absorbing plasmas are decoupled and have different location. Generally, both the Local Hot Bubble (LHB), the diffuse Galactic disk/halo and extragalactic background provide important and different contributions \citep[see e.g.][]{Kaastra2008, Wang2009}. The LHB is important at lower temperatures with a significant \ion{O}{vii} 1s-1p emission line at 0.57\,keV. Most of the observed \ion{O}{viii} emission belongs to the hot gas of the Galactic disk and halo, which is hotter than the LHB. Extragalactic sources provide the bulk of the X-ray background emission above 0.7\,keV. For instance, if we assume that the emitting gas has a density of $10^4$\,m$^{-3}$ (a common value in the local ISM) and $n_{\rm emi}/n_{\rm abs}\sim100$, we obtain that the emitting region has a depth of a few tens of parsec, close to the LHB. According to this scenario, most of the \ion{O}{vii} emission originates from the Local Hot Bubble, while the bulk of the absorption would be due to the diffuse hot interstellar gas present in the Galactic disk and halo. This means that our absorption measures are consistent with a Galactic origin for the interstellar clouds in the LOS towards Mrk~509 (and not WHIM) in agreement with previous UV work \citep{Savage2003, Sembach2003, Collins2004}.

We can probe the hot gas structure and location with a third alternative way. Following \citet{Yao09b} we predict the \ion{O}{vii-viii} line emission and column densities using their vertical exponential Galactic disk model. In their model, the gas density decays exponentially as a function of the height above the Galactic plane with a scale height of 2.8\,kpc and a central value of 1.4$\times10^{3}$\,m$^{-3}$, and the temperature has a scale height of 1.4\,kpc and a central value of 3.6$\times10^6$\,K. Their model parameters have large uncertainties, but can still provide useful constraints. Assuming a Galactic latitude of 30 degrees, on average this model predicts column densities of \ion{O}{vii}\,$=6^{+5}_{-2}\times10^{19}$\,m$^{-2}$ and \ion{O}{viii}\,$=2^{+3}_{-1}\times10^{19}$\,m$^{-2}$, originating from within a few kpc range from the Galactic plane. These column densities might contribute up to the 50\% of the absorbing hot gas in our LOS (see Table~\ref{table:empirical}). The remaining $\gtrsim$\,50\% of highly-ionized gas should belong to the {more distant} Galactic halo and the circumgalactic medium (CGM) and/or to the WHIM.

 \subsection{ISM column densities}
 \label{sec:discuss_column}

Most of the UV and X-ray interstellar lines in our spectra are not trivial to disentangle. The component groups HVCs (A, B, C), IVCs (D, F, G) and LVC~(E) provide smooth profiles, especially in the X-ray band where their blended profiles appear like a single line. In UV, the HVCs are well separated from the IVCs. However, each component produces only a handful of strong and not heavily saturated lines like those of \ion{Fe}{ii}, \ion{S}{ii} and \ion{C}{iv} (see Figs.~\ref{fig:cos_lines1}, \ref{fig:cos_lines2} and \ref{fig:cos_lines3}). The lines in the X-ray spectrum are not saturated, but due to the lower resolution the different velocity components form one blend. Moreover, both the emission and absorption lines intrinsic to the AGN partly affect the interstellar spectral range we fitted (see paper\,II).

\begin{table}
\renewcommand{\arraystretch}{1.3}
\caption{Column densities comparison for LVC component~E.}
\begin{center}
% use packages: array
 \small\addtolength{\tabcolsep}{+2pt}
\scalebox{1.1}{%
\begin{tabular}{l|lll}
\hline
      Par     &  Method 1$^{(b)}$ &  Method 2$^{(c)}$ & Method 3$^{(d)}$ \\
\hline
 \ion{N}{i}$\,^{(a)}$   & $19.5\pm0.2$   & $20.27\pm0.06$ & $20.44\pm0.06$   \\
 \ion{O}{i}$\,^{(a)}$   & $21.1\pm0.2$   & $21.4\pm0.1$   & $21.05\pm0.08$   \\
 \ion{Mg}{ii}$\,^{(a)}$ & $19.8\pm0.1$   & $<20.7$ & $19.83\pm0.05$   \\
 \ion{Fe}{ii}$\,^{(a)}$ & $19.3\pm0.1$   & $<19.8$ & $19.00\pm0.07$   \\
 \ion{O}{vi}$\,^{(a)}$  & $18.25\pm0.03$ & $18.9\pm0.3$   & $18.31\pm0.08$ \\
\hline
\end{tabular}}
\label{table:columns}
\end{center}
$^{(a)}$ The column densities $N_{\rm X}$ are in log (m$^{-2}$) units.\\
$^{(b)}$ Method 1 is the UV {\sl slab} model in Table~\ref{table:fit}.\\
$^{(c)}$ Method 2 refers to the X-ray {\sl slab} model in Table~\ref{table:empirical}.\\
$^{(d)}$ Method 3 gives the predictions of the physical model (Table~\ref{table:fit_physical}).
\end{table}

We have computed the ionic column densities predicted by the three different models (see Table~\ref{table:columns}). We compare only the results obtained for component E, as this component provides the best constrained column densities and actually they are the only free parameters in the RGS fits. The reason for using an empirical model for both the UV and X-ray spectra is to identify discrepancies due to saturation between the UV and X-ray determined ionic column densities. The UV (Method~1) and X-ray (Method~2) column densities are mostly consistent (see Table~\ref{table:columns}). Higher values for \ion{N}{i} and \ion{O}{vi} column densities are obtained from the RGS spectrum. The column in UV for \ion{N}{i} is clearly under-estimated because of saturation as confirmed by the different line ratios of the components E and F in the 1200\,{\AA} triplet (see Figs.~\ref{fig:cos_lines1} and \ref{fig:cold_gas}). The \ion{O}{vi} column density discrepancy is smaller, about 2$\sigma$. The difference is most likely due to the weakness of the \ion{O}{vi} 1s-2p line at 22\,{\AA} and the spectral noise (see Fig.~\ref{fig:full_model}). Apart from the already discussed \ion{N}{i}, the UV/X-ray model (Method~3) provides column densities consistent with the UV {\sl slab} model.

 \subsection{ISM abundances}
 \label{sec:discuss_abundances}

\begin{table}
\renewcommand{\arraystretch}{1.}
\caption{Total abundances for the cold and warm phases of IVCs.}
\begin{center}
% use packages: array
 \small\addtolength{\tabcolsep}{+0pt}
\scalebox{1.}{%
\begin{tabular}{l|lll}
\hline
                   & Cold gas      &  Cold phase$^{(b)}$ &  Warm phase           \\
\hline
  C/H$\,^{(a)}$    & $<0.1$        & $0.7\pm0.1$         &  $\equiv 1$           \\
  N/H$\,^{(a,c)}$  & $0.8\pm0.1$   & $0.8\pm0.1$         &  $0.8\pm0.1$          \\
  O/H$\,^{(a)}$    & $0.5\pm0.1$   & $0.7\pm0.1$         &  $0.5\pm0.1\,(+0.4)^{(e)}$  \\
  Ne/H$\,^{(a,c)}$ & $1.3\pm0.1$   & $1.3\pm0.1$         &  $1.3\pm0.1$          \\
  Mg/H$\,^{(a,c)}$ & $0.41\pm0.04$ & $\leq1.0\,^{(d)}$   &  $0.4-1.0$            \\
  Al/H$\,^{(a)}$   & $0.06\pm0.01$ & $\leq1.3\,^{(d)}$   &  $0.33\pm0.08\,(+0.3)^{(e)}$\\
  Si/H$\,^{(a)}$   & $0.14\pm0.04$ & $\leq0.62^{(d)}$    &  $0.60\pm0.01$        \\
  S/H$\,^{(a,c)}$  & $0.67\pm0.01$ & $0.67\pm0.01$       &  $0.67\pm0.01$        \\
  Fe/H$\,^{(a)}$   & $0.07\pm0.01$ & $\leq1.0\,^{(d)}$   &  $0.90\pm0.06$        \\
  Ni/H$\,^{(a,c)}$ & $0.07\pm0.01$ & $\leq1.8\,^{(d)}$   &  $0.7-1.8$            \\
\hline
\end{tabular}}
\label{table:abundances}
\end{center}
$^{(a)}$ Abundances ratios in proto-Solar units of \citet{Lodders09}.\\
$^{(b)}$ The cold phase includes cold gas and molecules from Table~\ref{table:fit_physical}.\\
$^{(c)}$ Cold and warm phase abundances are coupled.\\
$^{(d)}$ Molecules and dust predicted upper limits.\\
$^{(e)}$ Systematic errors (see also Sect.~\ref{sec:discuss_abundances}).
\end{table}

The ISM abundances are computed by taking into account the contribution from gas, dust and molecules. A thorough and accurate analysis can be done only for the rest frame component, which is detected in all gas and molecular phases (see Table~\ref{table:fit_physical}). We sum the gas and dust contributions to the cold phase for component E, calculate the total abundances and compare them to those of the warm phase in Table~\ref{table:abundances}.

There are some limitations to determining the abundances, for instance the non-detection of aluminates allows us only to give the gas contribution to the aluminum abundance. Aluminates absorption features are weak and a fit of the O~K edge with additional FeAl$_2$O$_4$ just provides an upper limit to the aluminum abundance of about 1.3 in units of \cite{Lodders09}, see Table~\ref{table:abundances}. The CO column density can be estimated, while for the other molecules only upper limits are obtained. Thus, we can give only the predicted upper limits to the abundances of Mg, Si, and Fe. Moreover, Mg cannot be detected in the warm phase because most \ion{Mg}{ii} is already provided by the cold gas, and intermediate-ionization ions do not provide relevant absorption lines in our energy domain, thus its abundance has been coupled to that of the cold gas (see also Sect.~\ref{sec:Probe_abundances}). The same applies to Ni, for we have tested a fit with additional NiO molecules, which have provided an upper limit to the nickel abundance of about 1.8. As mentioned above the abundances of N, Ne and S have been coupled between the two phases as those elements are not suspected to be depleted from the cold phase into dust grains.

A comparison between the cold and the warm phases is worthwhile despite these limitations. First, we find consistent abundances for O, Si and Fe for both phases once dust is accounted for. The total iron and nitrogen abundance is close to the proto-Solar value. Differently, $\alpha$ elements like O, Al, Si and S appear to be under-abundant. The Si and S abundance measurements have a high confidence due to the several lines that have been used. The Al and O abundances might be under-estimated. The Al column density has been measured with the 1670.8\,{\AA} absorption line which is heavily saturated. If we fit the Al absorption line by ignoring its saturated bottom, we get a systematic error on the abundance of about 0.3, i.e. about 100\% (see Table~\ref{table:abundances}). The oxygen low-ionization lines are highly affected by the AGN intrinsic features. Indeed, within 1$\sigma$ error of the AGN model (see paper\,III) we obtain a systematic error on the ISM oxygen abundance of 0.4 for the warm gas. We might also miss some additional contribution to the ISM oxygen from molecules as previously discussed (see Sect.~\ref{sec:disc_cold_phase}). The carbon abundance in the cold phase appears to be lower than in the warm phase, which likely means that we miss some additional molecules like hydrocarbons. Models predict an optical depth of about 0.1 at 43\,{\AA}, but here the S/N ratio is too low even for the LETGS spectrum in order to measure column densities (see paper\,V).  Neon is the only over-abundant element, which might suggest that the proto-Solar Ne abundance we use is under-estimated \citep[see e.g.][]{DrakeTesta, Pinto2010}.

Unfortunately statistics are not high enough to estimate the abundances of the hot ionized gas because its lines are much weaker than those of the cold and warm gas. We note, however, that the high-ionization phase can be well fitted by adopting proto-Solar abundances.

\subsection{The characteristics of LVC, IVCs and HVCs}
\label{sec:ISM_forest}

\begin{figure}
      \centering
      \includegraphics[bb=70 80 515 710, angle=90, width=9cm]{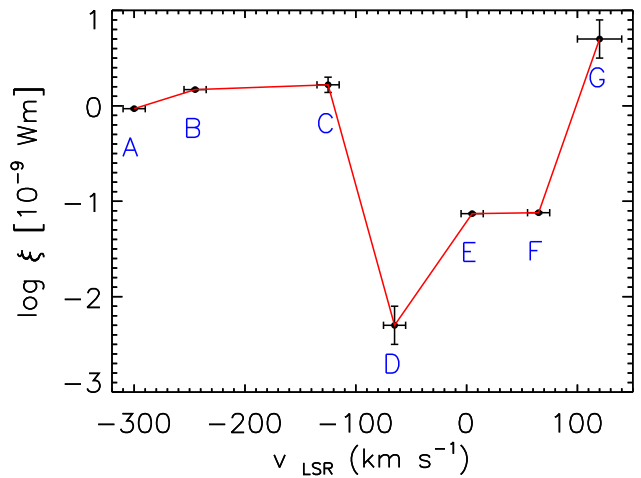}%}
      \caption{Ionization parameter versus average velocity for the ISM warm phase of the seven cloud systems (see also Table~\ref{table:fit_physical}).}
          \label{fig:xi_vs_v}
    \end{figure}

The separation between IVCs and HVCs is justified if we compare the properties of the two groups (see Table~\ref{table:fit_physical}). The HVCs are clouds highly ionized by both UV\,/\,X-ray background photons and collisions {with the surrounding hot gas}. Neutral gas is absent at these high speeds in this LOS. The HVCs show proto-Solar Si\,/\,C ratio, while the IVCs have a carbon excess. Moreover, the ionization parameters $\xi$ of the HVC warm gas are higher than those of the IVCs (see also Fig.~\ref{fig:xi_vs_v}). If the clouds are effectively photo-ionized by the same QSO\,/\,Galactic SED, consistent with our best-fit results, this might suggest that the slow components E and F are in a region closer to the Galactic disk, where less UV and X-ray photons penetrate.

 \subsubsection{LVC location: metallicity method}
 \label{sec:LVC_location}

In order to locate the interstellar absorbers we can compare our measurement of the iron abundance with the Galactic metallicity gradient \cite[see also][]{Pinto2010}. We use the abundance for the warm phase of component E as it is well constrained. We adopt a Galactic altitude of zero for the Sun as it is less than 30~pc distant from the Galactic plane, which is much smaller than the kpc scales we are interested in. If $A$ is the abundance of element $X$ in an interstellar cloud, $r$ the difference between the Galactocentric radii of the cloud and that of the Sun, and $h$ the cloud altitude, then the abundance change in the line of sight can be written as
\begin{equation}
\Delta\,A = r\,\frac{dA}{dr}+\frac{dA}{dh}\,h.
\label{eq:abundances1}
\end{equation}
The radial and vertical dependence of the abundance can be expressed as $A_r = A_{\odot}\,10^{\,\alpha_r\,r}$ and $A_h = A_{\odot}\,10^{\,\alpha_h\,h}$ \citep[see e.g.][]{GradPedicelli}, where $\alpha_r$ and $\alpha_h$ are the radial and vertical slopes of the abundance gradient, and $A_{\odot}$ is the proto-Solar value \citep{Lodders09}. We estimate the slopes by calculating the average metallicity values found in the literature \citep{Yamagata1994, Maciel2009, GradPedicelli, Chen2011}. We find $\alpha_r\sim-0.06$ and $\alpha_h\sim-0.11$ for iron, which means that the vertical gradient is steeper than the radial one. From Eq.~(\ref{eq:abundances1}) we obtain
\begin{equation}
\frac{A}{A_{\odot}} = \left (\alpha_r\,r \cdot 10^{\,\alpha_r\,r} + \alpha_h\,h \cdot 10^{\,\alpha_h\,h} \right ) \cdot \ln\,10 + 1.
\label{eq:abundances2}
\end{equation}
Through Eq.~(\ref{eq:abundances2}) we calculate the Fe abundances for a grid of distances $d$ starting from 0.1\,kpc and compare them with our Fe abundance measurement for the warm gas of component~E (see Table~\ref{table:abundances}). This provides $d\,\leq\,0.5$\,kpc for component~E, which means that the bulk of the warm interstellar absorption is local and belongs to the Galactic disk. The cold gas exactly follows the kinematics of the warm gas and thus is co-located in the Galactic disk as well.

\subsubsection{LVC and IVCs location: kinematics method}
\label{sec:IVCs_location}

\begin{figure}
      \centering
      \includegraphics[bb=70 80 515 710, angle=90, width=9cm]{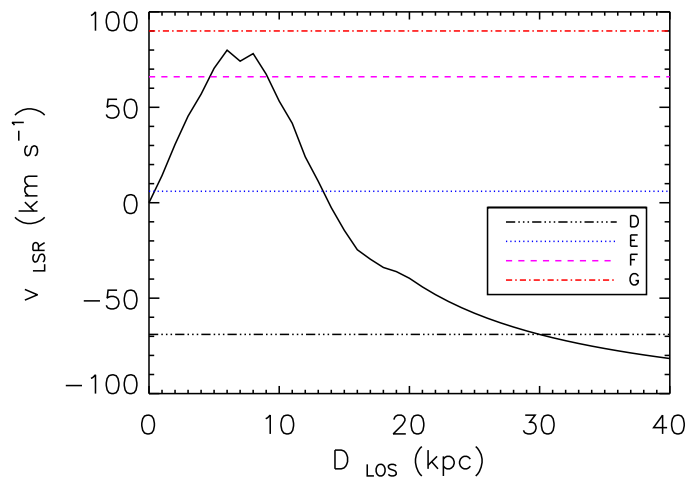}%}
      \caption{Rotation curve of the Galaxy in the LOS towards Mrk~509 (see Sect.~\ref{sec:IVCs_location}). The velocities of the LVC and IVCs are also displayed.}
          \label{fig:rotation}
    \end{figure}

It is possible to probe the location of the LVC and IVCs by comparing their LSR velocities with the Galactic rotation. For HVCs this is not possible as their LSR velocities are outside the observed range for Galactic rotation. In Fig.~\ref{fig:rotation} we show the rotation curve of the Galaxy as function of the distance as measured in the LOS towards Mrk~509 \citep[see chapter~9 of][]{Binney1998}.

Component E, with $v_{\rm LSR}=0-10$\,km\,s$^{-1}$ (see Table~\ref{table:fit}), is consistent with distances of $0-1$\,kpc or $13-15$\,kpc. However, we note that the density strongly decreases with the height from the Galactic plane. At a distance of 14\,kpc along the LOS of Mrk~509, the height above the Galactic plane is 7 kpc, where there is much less neutral gas as within a few hundred pc from the Galactic plane. This would imply that component~E is thus $<$\,1\,kpc far away, consistent with our earlier conclusion in Sect.~\ref{sec:LVC_location}.

Component F shows velocities of $60-70$\,km\,s$^{-1}$ and for the same reason, namely the decrease in density, it must be at a distance of $4-6$\,kpc. The large distance from both the Earth and the Galactic plane is confirmed by the smaller neutral column densities of component F, which are lower than those of component E by at least an order of magnitude (see Table~\ref{table:fit}). The small distance between component E and F also explains why a very good fit is obtained by adopting the similar abundances, temperatures and ionization parameters found for both clouds (Table~\ref{table:fit_physical}).

The location of the remaining two IVCs is not as easy to determine because the uncertainties are larger. If component D ($v_{\rm LSR}\sim-65$\,km\,s$^{-1}$) belongs to the Galactic corotating gas, then it must have a distance of $25-30$\,kpc (residing at the other side of the Galaxy) and a height from the plane larger than $12$\,kpc, which explains why its column densities are two orders of magnitude lower than those for component E.

It is more difficult to determine the position of component G. Fitting the different ions has shown a spread of about $40$\,km\,s$^{-1}$ in the measured velocities. \ion{Fe}{ii} and \ion{S}{ii} are consistent with $v_{\rm LSR}=90$\,km\,s$^{-1}$, while most ions show velocities near $130$\,km\,s$^{-1}$ (see Table~\ref{table:fit}). The spread is larger than the COS wavelength calibration uncertainties and seems to suggest a double nature of component G. The smallest contribution might refer to corotating gas moving at $\sim80-90$\,km\,s$^{-1}$ with a distance of $6-8$\,kpc. The bulk of component G is represented by the warm phase and moving with a velocity consistent in modulus with those of the HVCs. Unfortunately, the signal-to-noise ratio is small for component G and the $\chi^2_{\nu}$ does not significantly change by dividing this component into two subcomponents, one at $90$ and the other at $130$\,km\,s$^{-1}$. If the fast component refers to a Galactic fountain, then it should be close to the Galactic plane as these fountains are expected to reach heights of at most a few kpc. In fact, the high temperatures measured for the cold and warm gas of component G might be signatures of shocks due to interaction with its surrounding (see Table~\ref{table:fit_physical}). An alternative explanation for component G is that it is part of a captured extragalactic cloud, although a satisfactory fit is obtained assuming the same abundances as for Galactic components D, E, and F.

\subsubsection{HVCs location: ionization structure}
\label{sec:HVCs_location}

\begin{figure}
      \centering
      \includegraphics[angle=0, width=9cm]{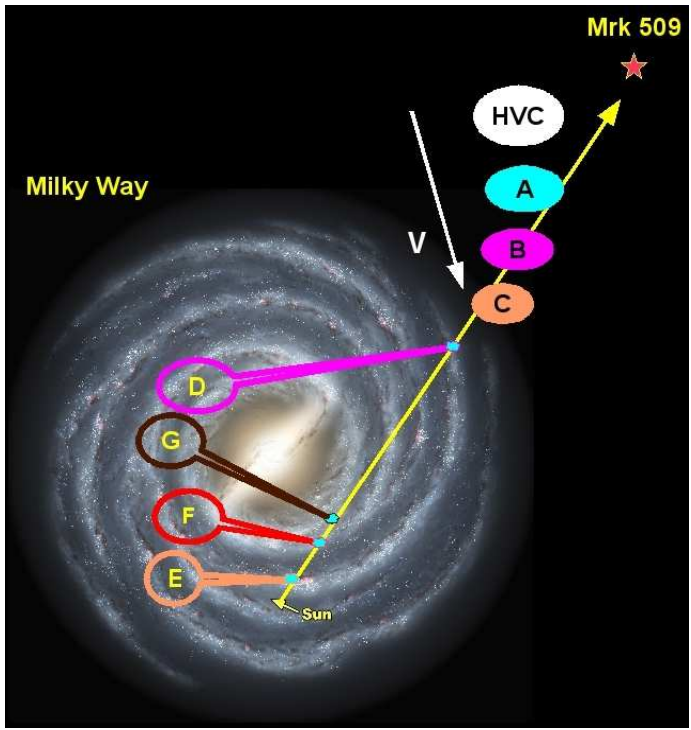}%}
      \caption{Artistic view of the IVCs (components D--G) laying near the Galactic disk and HVCs (A--C) impacting the Galactic halo.}
          \label{fig:HVCs}
    \end{figure}

A different approach is required to determine the location of the three HVCs (components A, B and C). All three components have been successfully modeled by adopting the same abundances and photo-ionizing source, extra collisional ionization, and a high ionization state. Interestingly, no sign of fast neutral gas was detected along this LOS, but \citet{Sembach1999} found evidence for \ion{H}{i} 21\,cm emission at these velocities at a distance of about 2 degrees from Mrk~509. One plausible scenario is that the three ionized HVCs are the outskirt of one large captured cloud, maybe matter stripped from a satellite galaxy, which has fallen into the gravitational well of the Milky Way (see also Fig.~\ref{fig:HVCs}). The hot collisionally-ionized gas suggests that the cloud is impacting the halo of the Galaxy or the Local Group. The fastest cloud, component A, is less ionized than the other 2 HVCs, possibly because it is more distant and thus has currently a smaller interaction with the Galactic halo. This implies that the interaction causes the infalling gas to slowdown.

\subsubsection{Equilibrium in the interstellar clouds forest}
\label{sec:equilibrium}

\begin{figure}
      \centering
      \includegraphics[bb=70 80 515 680, angle=90, width=9cm]{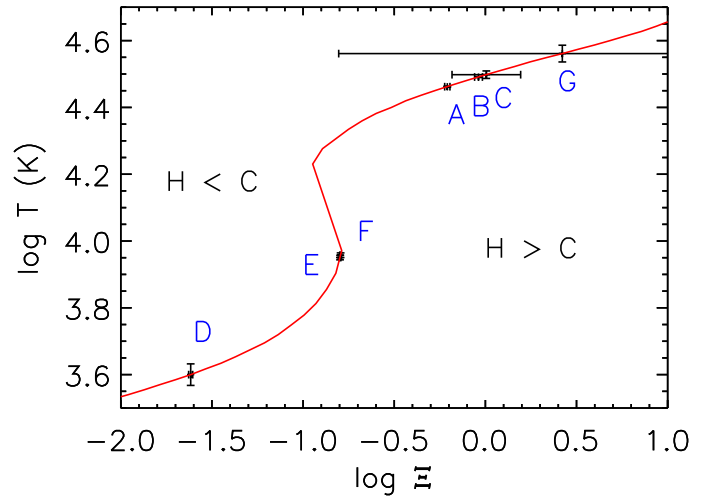}%}
      \caption{Stability curve for the LE SED. The values for all the warm photo-ionized components A--G are also displayed (see Sects.~\ref{sec:Probe_warm_gas} and \ref{sec:equilibrium}).}
          \label{fig:bcurve}
    \end{figure}

Here we propose the stability curve as an alternative method to probe the dynamical structure of the ISM and to test whether some of the components are co-located. This curve shows the relation between the temperature $T$ of the gas and its pressure $\Xi$, which is commonly defined as
\begin{equation}
\Xi = \frac{L}{4 \pi r^2 c p} = \frac{\xi}{6 \pi c k T}.
\label{eq:pressure}
\end{equation}
This curve was calculated using the SPEX {\sl xabsinput} tool in the computation of the ionization balance of the photo-ionized warm phase (see Sect.~\ref{sec:Probe_warm_gas}, LE SED) and it is displayed in Fig.~\ref{fig:bcurve} together with the values calculated for components A--G. This curve divides the $T-\Xi$ plane in two regions: in the region below the curve the heating dominates the cooling, while above the curve the cooling dominates the heating \citep{Krolik1981}. The curve segments which have positive slopes are stable against thermal perturbations, while those with a negative slope are not. Moreover components which have the same $\Xi$ value on this curve are in pressure equilibrium and are thus likely part of the same interstellar structure.

Apparently, all the seven clouds are on stable branches of the curve (see Fig.~\ref{fig:bcurve}). Apart from component~G, which is not well constrained, the LVC~(E), IVCs (D, F and G) and HVCs (A, B and C) show a rather different $\Xi$ value. As expected the IVC and HVC groups have a different nature and are not co-located. Components E and F share the same $\Xi$ and are thus supposed to belong to the Galactic disk environment as we have previously shown. Component~D differs from all the others and should not coexist with them. As expected, HVC components A, B and C have quite similar $\Xi$, which means that they might belong to the same structure. The uncertainty on the $\Xi$ value, due to the large error in the $\xi$ value, makes the results for component~G uncertain.

 \subsection{Comparison with previous results}
 \label{sec:discuss_compare}

Although the interstellar clouds show a different nature and origin in the LOS towards Mrk~509, we have proved that the ISM structure is well described by three phases with different ionization states (see Sect.~\ref{sec:complete_model}). The cold phase is a blend of molecules, dust and low-ionization gas, and is not detected in the HVCs. The warm phase is characterized by mildly ionized gas in photo-ionization equilibrium and is present in both HVCs and IVCs. The hot phase is characterized by collisional ionization. These results agree well with the current state of the art \citep{ferriere} as well as with what was found by \citet{Pinto2010} in the LOS towards \object{GS 1826--238}, which is a LMXB located near the Galactic Center. This means that the ISM of the Galaxy follows a certain equilibrium on Galactic scales due to its cooling and heating processes.

We have measured the column densities of the different phases for all the seven cloud systems (see Table~\ref{table:fit}) and compared with those found in the literature. Within the errors, our estimates agree well with the column densities found by previous work \citep{Sembach1995, Sembach1999, Sembach2003, Collins2004, Shull2009}. Our \ion{O}{vi} column densities for the HVCs A--C and the IVC~G agree within $1\sigma$ with those measured by \citet{Fox2006}, who take into account the H$_2$ absorption, which affects this wavelength range (see Fig.~\ref{fig:warm_gas}). They consider the blend of components A and B as one absorber, but their column density estimate matches the sum of the values that we have obtained separately for components A and B. Our total \ion{H}{i} column density differs from the value measured at long wavelengths (see Sect.~\ref{sec:Probe_cold_gas}). This is most likely due to the difference between the beam-size of the UV and radio observations.

Our determination of the location of the different clouds agrees with previous work, but improves upon the previous results. As discussed in Sect.~\ref{sec:HVCs_location}, the HVCs (A--C) are likely at the outskirts of an extragalactic cloud captured by the Galaxy or the Local Group and interacting with them as suggested by \citet{Sembach1999}. {Component~B has a $|v|$ lower than component~A maybe due to a stronger impact with the Galactic hot gas, as suggested by its higher ionization state. This might indicate that the external layers of the cloud slewdown} during the infall, in agreement with the scenario proposed by \citet{Lehner2011}. Such clouds, known also as VHVCs (see Sect.~\ref{sec:introduction}), are thought to lose their \ion{H}{i} and a significant amount of speed during the infall and interaction with the Galaxy. Moreover, from the similar total pressure and temperature properties measured for the warm phase, we conclude that they likely belong to the same cloud structure (see Fig.~\ref{fig:bcurve}). Components~E ($\sim0$\,km\,s$^{-1}$) and F ($\sim60$\,km\,s$^{-1}$) have very similar properties for their cold and warm phases and clearly correspond to interstellar gas which follows the Galactic rotation and are expected to reside at less than 1 and 5\,kpc, respectively (see Fig.~\ref{fig:rotation} and Sect.~\ref{sec:discuss_abundances}). This is fully consistent with the conclusions by \citet{Blades1983}. Component~D ($\sim-60$\,km\,s$^{-1}$) is most likely located off the Galactic plane as previously suggested by \citet{Morton1986}, but should be located at high altitudes.

\section{Conclusion}
\label{sec:conclusion}

We have presented a complete analysis of the interstellar and circumgalactic medium towards the AGN Mrk~509, which is a bright X-ray source with a high Galactic latitude of about $-30^\circ$, {through high-quality grating spectra taken with \textit{XMM-Newton}~/~RGS, HST~/~COS and FUSE.}
\begin{itemize}
 \item On average the ISM, as found in the form of LVCs and IVCs near the Galactic disk, preserves a structure consisting of three distinct main phases with different ionization states. HVCs, which usually probe the halo and CGM, have a different structure and in the LOS towards Mrk~509 they show two main phases.
 \item We have probed seven different cloud systems along the LOS. Our study of their kinematics and abundances has revealed the nature and location of the clouds with respect to the Galactic environment assuming Galactic rotation.
 \item The HVCs (components A--C) are highly ionized most likely as the result of a larger exposure to the X-ray background and the interaction with the circumgalactic medium. The similar abundances and ionization structure suggest a common origin for these HVCs. They {might be} at the outskirt of an extragalactic cloud captured by the Galaxy.
 \item The LVC (component~E) and IVC (component~F) refer to interstellar gas co-rotating with the Galactic disk, as confirmed by the detection of dust and molecules at rest wavelengths.
 \item The location of the IVCs (components D and G) is not well constrained, they might belong either to disk co-rotating gas or to Galactic fountains.
 \item The column densities, ionization states and locations probed by our alternative models/methods agree with each other and with the results in the literature.
All this supports our research method and justifies extension to more lines-of-sight and/or new observations. These analyses will provide indeed a better mapping of the ISM and a deeper study of its chemical composition and interaction with the entire Galaxy.
\end{itemize}

\begin{acknowledgements}
This work is based on observations obtained with XMM-Newton, an
ESA science mission with instruments and contributions directly funded by
ESA Member States and the USA (NASA). SRON is supported financially by
NWO, the Netherlands Organization for Scientific Research. We acknowledge support
from NASA/XMM-Newton Guest Investigator grant NNX09AR01G. Support for HST Program number
12022 was provided by NASA through grants from the Space Telescope Science
Institute, which is operated by the Association of Universities for Research
in Astronomy, Inc., under NASA contract NAS5-26555. We also acknowledge Rob Detmers 
for providing us with the AGN continuum model of Mrk~509. We thank Ken Sembach for
helpful discussions on Galactic halo gas and HVCs. We also acknowledge the referee
for the useful suggestions provided to us, which improve the quality of the paper.
\end{acknowledgements}

\bibliographystyle{aa}
\bibliography{bibliografia} %----> bibliografia.bib

\begin{thebibliography}{52}
\expandafter\ifx\csname natexlab\endcsname\relax\def\natexlab#1{#1}\fi

\bibitem[{{Binney} \& {Merrifield}(1998)}]{Binney1998}
{Binney}, J. \& {Merrifield}, M. 1998, {Galactic Astronomy}, ed. {Binney, J.~\&
  Merrifield, M.}

\bibitem[{{Blades} \& {Morton}(1983)}]{Blades1983}
{Blades}, J.~C. \& {Morton}, D.~C. 1983, \mnras, 204, 317

\bibitem[{{Blitz} {et~al.}(1999){Blitz}, {Spergel}, {Teuben}, {Hartmann}, \&
  {Burton}}]{Blitz1999}
{Blitz}, L., {Spergel}, D.~N., {Teuben}, P.~J., {Hartmann}, D., \& {Burton},
  W.~B. 1999, \apj, 514, 818

\bibitem[{{Chakravorty} {et~al.}(2009){Chakravorty}, {Kembhavi}, {Elvis}, \&
  {Ferland}}]{Chakravorty09}
{Chakravorty}, S., {Kembhavi}, A.~K., {Elvis}, M., \& {Ferland}, G. 2009,
  \mnras, 393, 83

\bibitem[{{Chen} {et~al.}(2011){Chen}, {Zhao}, {Carrell}, \& {Zhao}}]{Chen2011}
{Chen}, Y.~Q., {Zhao}, G., {Carrell}, K., \& {Zhao}, J.~K. 2011, \aj, 142, 184

\bibitem[{{Collins} {et~al.}(2004){Collins}, {Shull}, \&
  {Giroux}}]{Collins2004}
{Collins}, J.~A., {Shull}, J.~M., \& {Giroux}, M.~L. 2004, \apj, 605, 216

\bibitem[{{Costantini} {et~al.}(2012){Costantini}, {Pinto}, {Kaastra}, {in't
  Zand}, {Freyberg}, {Kuiper}, {M{\'e}ndez}, {de Vries}, \&
  {Waters}}]{Costantini2012}
{Costantini}, E., {Pinto}, C., {Kaastra}, J.~S., {et~al.} 2012, \aap, 539, A32

\bibitem[{{den Herder} {et~al.}(2001){den Herder}, {Brinkman}, {Kahn},
  {Branduardi-Raymont}, {Thomsen}, {Aarts}, {Audard}, {Bixler}, {den Boggende},
  {Cottam}, {Decker}, {Dubbeldam}, {Erd}, {Goulooze}, {G{\"u}del}, {Guttridge},
  {Hailey}, {Janabi}, {Kaastra}, {de Korte}, {van Leeuwen}, {Mauche},
  {McCalden}, {Mewe}, {Naber}, {Paerels}, {Peterson}, {Rasmussen}, {Rees},
  {Sakelliou}, {Sako}, {Spodek}, {Stern}, {Tamura}, {Tandy}, {de Vries},
  {Welch}, \& {Zehnder}}]{denherder01}
{den Herder}, J.~W., {Brinkman}, A.~C., {Kahn}, S.~M., {et~al.} 2001, \aap,
  365, L7

\bibitem[{{Detmers} {et~al.}(2011){Detmers}, {Kaastra}, {Steenbrugge},
  {Ebrero}, {Kriss}, {Arav}, {Behar}, {Costantini}, {Branduardi-Raymont},
  {Mehdipour}, {Bianchi}, {Cappi}, {Petrucci}, {Ponti}, {Pinto}, {Ratti}, \&
  {Holczer}}]{Detmers11}
{Detmers}, R.~G., {Kaastra}, J.~S., {Steenbrugge}, K.~C., {et~al.} 2011, \aap,
  534, A38, {(paper III)}

\bibitem[{{Drake} \& {Testa}(2005)}]{DrakeTesta}
{Drake}, J.~J. \& {Testa}, P. 2005, \nat, 436, 525

\bibitem[{{Ebrero} {et~al.}(2011){Ebrero}, {Kriss}, {Kaastra}, {Detmers},
  {Steenbrugge}, {Costantini}, {Arav}, {Bianchi}, {Cappi},
  {Branduardi-Raymont}, {Mehdipour}, {Petrucci}, {Pinto}, \&
  {Ponti}}]{Ebrero11}
{Ebrero}, J., {Kriss}, G.~A., {Kaastra}, J.~S., {et~al.} 2011, \aap, 534, A40,
  {(paper V)}

\bibitem[{{Ferland}(2005)}]{Ferland2005}
{Ferland}, G.~J. 2005, {Hazy, A Brief Introduction to Cloudy 05.07}, ed.
  {Ferland, G.~J.}

\bibitem[{{Ferri{\`e}re}(2001)}]{ferriere}
{Ferri{\`e}re}, K.~M. 2001, Reviews of Modern Physics, 73, 1031

\bibitem[{{Fox} {et~al.}(2006){Fox}, {Savage}, \& {Wakker}}]{Fox2006}
{Fox}, A.~J., {Savage}, B.~D., \& {Wakker}, B.~P. 2006, \apjs, 165, 229

\bibitem[{{Ghavamian} {et~al.}(2009){Ghavamian}, {Aloisi}, {Lennon}, {Hartig},
  {Kriss}, {Oliveira}, {Massa}, {Keyes}, {Proffitt}, {Delker}, \&
  {Osterman}}]{Ghavamian09}
{Ghavamian}, P., {Aloisi}, A., {Lennon}, D., {et~al.} 2009, {Preliminary
  Characterization of the Post- Launch Line Spread Function of COS}, Tech. rep.

\bibitem[{{Gnat} \& {Sternberg}(2007)}]{Gnat2007}
{Gnat}, O. \& {Sternberg}, A. 2007, \apjs, 168, 213

\bibitem[{{Green} {et~al.}(2012){Green}, {Froning}, {Osterman}, {Ebbets},
  {Heap}, {Leitherer}, {Linsky}, {Savage}, {Sembach}, {Shull}, {Siegmund},
  {Snow}, {Spencer}, {Stern}, {Stocke}, {Welsh}, {B{\'e}land}, {Burgh},
  {Danforth}, {France}, {Keeney}, {McPhate}, {Penton}, {Andrews},
  {Brownsberger}, {Morse}, \& {Wilkinson}}]{Green2011}
{Green}, J.~C., {Froning}, C.~S., {Osterman}, S., {et~al.} 2012, \apj, 744, 60

\bibitem[{{Haardt} \& {Madau}(2001)}]{Haardt2001}
{Haardt}, F. \& {Madau}, P. 2001, in Clusters of Galaxies and the High Redshift
  Universe Observed in X-rays, ed. {D.~M.~Neumann \& J.~T.~V.~Tran}

\bibitem[{{Haardt} \& {Madau}(2012)}]{Haardt2011}
{Haardt}, F. \& {Madau}, P. 2012, \apj, 746, 125

\bibitem[{{Kaastra} {et~al.}(2009){Kaastra}, {de Vries}, {Costantini}, \& {den
  Herder}}]{kaastra09}
{Kaastra}, J.~S., {de Vries}, C.~P., {Costantini}, E., \& {den Herder},
  J.~W.~A. 2009, \aap, 497, 291

\bibitem[{{Kaastra} {et~al.}(2011{\natexlab{a}}){Kaastra}, {de Vries},
  {Steenbrugge}, {Detmers}, {Ebrero}, {Behar}, {Bianchi}, {Costantini},
  {Kriss}, {Mehdipour}, {Paltani}, {Petrucci}, {Pinto}, \&
  {Ponti}}]{Kaastra11b}
{Kaastra}, J.~S., {de Vries}, C.~P., {Steenbrugge}, K.~C., {et~al.}
  2011{\natexlab{a}}, \aap, 534, A37, {(paper II)}

\bibitem[{{Kaastra} {et~al.}(1996){Kaastra}, {Mewe}, \&
  {Nieuwenhuijzen}}]{kaastraspex}
{Kaastra}, J.~S., {Mewe}, R., \& {Nieuwenhuijzen}, H. 1996, in UV and X-ray
  Spectroscopy of Astrophysical and Laboratory Plasmas, ed. {K.~Yamashita \&
  T.~Watanabe}, 411

\bibitem[{{Kaastra} {et~al.}(2008){Kaastra}, {Paerels}, {Durret}, {Schindler},
  \& {Richter}}]{Kaastra2008}
{Kaastra}, J.~S., {Paerels}, F.~B.~S., {Durret}, F., {Schindler}, S., \&
  {Richter}, P. 2008, \ssr, 134, 155

\bibitem[{{Kaastra} {et~al.}(2011{\natexlab{b}}){Kaastra}, {Petrucci}, {Cappi},
  {Arav}, {Behar}, {Bianchi}, {Bloom}, {Blustin}, {Branduardi-Raymont},
  {Costantini}, {Dadina}, {Detmers}, {Ebrero}, {Jonker}, {Klein}, {Kriss},
  {Lubi{\'n}ski}, {Malzac}, {Mehdipour}, {Paltani}, {Pinto}, {Ponti}, {Ratti},
  {Smith}, {Steenbrugge}, \& {de Vries}}]{Kaastra11a}
{Kaastra}, J.~S., {Petrucci}, P.-O., {Cappi}, M., {et~al.} 2011{\natexlab{b}},
  \aap, 534, A36, {(paper I)}

\bibitem[{{Kriss}(2011)}]{Kriss11a}
{Kriss}, G.~A. 2011, {Improved Medium Resolution Line Spread Functions for COS
  FUV Spectra}, Tech. rep.

\bibitem[{{Kriss} {et~al.}(2011){Kriss}, {Arav}, {Kaastra}, {Ebrero}, {Pinto},
  {Borguet}, {Edmonds}, {Costantini}, {Steenbrugge}, {Detmers}, {Behar},
  {Bianchi}, {Blustin}, {Branduardi-Raymont}, {Cappi}, {Mehdipour}, {Petrucci},
  \& {Ponti}}]{Kriss11b}
{Kriss}, G.~A., {Arav}, N., {Kaastra}, J.~S., {et~al.} 2011, \aap, 534, A41,
  {(paper VI)}

\bibitem[{{Kriss} {et~al.}(2000){Kriss}, {Green}, {Brotherton}, {Oegerle},
  {Sembach}, {Davidsen}, {Friedman}, {Kaiser}, {Zheng}, {Woodgate},
  {Hutchings}, {Shull}, \& {York}}]{Kriss2000}
{Kriss}, G.~A., {Green}, R.~F., {Brotherton}, M., {et~al.} 2000, \apjl, 538,
  L17

\bibitem[{{Krolik} {et~al.}(1981){Krolik}, {McKee}, \& {Tarter}}]{Krolik1981}
{Krolik}, J.~H., {McKee}, C.~F., \& {Tarter}, C.~B. 1981, \apj, 249, 422

\bibitem[{{Lehner} \& {Howk}(2011)}]{Lehner2011}
{Lehner}, N. \& {Howk}, J.~C. 2011, Science, 334, 955

\bibitem[{{Lodders} \& {Palme}(2009)}]{Lodders09}
{Lodders}, K. \& {Palme}, H. 2009, Meteoritics and Planetary Science
  Supplement, 72, 5154

\bibitem[{{Maciel} \& {Costa}(2009)}]{Maciel2009}
{Maciel}, W.~J. \& {Costa}, R.~D.~D. 2009, in IAU Symposium, Vol. 254, IAU
  Symposium, ed. {J.~Andersen, J.~Bland-Hawthorn, \& B.~Nordstr{\"o}m}, 38P

\bibitem[{{McGee} \& {Newton}(1986)}]{McGee1986}
{McGee}, R.~X. \& {Newton}, L.~M. 1986, Proceedings of the Astronomical Society
  of Australia, 6, 358

\bibitem[{{Morton} \& {Blades}(1986)}]{Morton1986}
{Morton}, D.~C. \& {Blades}, J.~C. 1986, \mnras, 220, 927

\bibitem[{{Murphy} {et~al.}(1996){Murphy}, {Lockman}, {Laor}, \&
  {Elvis}}]{Murphy1996}
{Murphy}, E.~M., {Lockman}, F.~J., {Laor}, A., \& {Elvis}, M. 1996, \apjs, 105,
  369

\bibitem[{{Netzer}(2008)}]{Netzer2008}
{Netzer}, H. 2008, \nar, 52, 257

\bibitem[{{Pedicelli} {et~al.}(2009){Pedicelli}, {Bono}, {Lemasle}, {Fran{\c
  c}ois}, {Groenewegen}, {Lub}, {Pel}, {Laney}, {Piersimoni}, {Romaniello},
  {Buonanno}, {Caputo}, {Cassisi}, {Castelli}, {Leurini}, {Pietrinferni},
  {Primas}, \& {Pritchard}}]{GradPedicelli}
{Pedicelli}, S., {Bono}, G., {Lemasle}, B., {et~al.} 2009, \aap, 504, 81

\bibitem[{{Pinto} {et~al.}(2010){Pinto}, {Kaastra}, {Costantini}, \&
  {Verbunt}}]{Pinto2010}
{Pinto}, C., {Kaastra}, J.~S., {Costantini}, E., \& {Verbunt}, F. 2010, \aap,
  521, A79

\bibitem[{{Richter}(2006)}]{Richter2006}
{Richter}, P. 2006, in Reviews in Modern Astronomy, Vol.~19, Reviews in Modern
  Astronomy, ed. {S.~Roeser}, 31

\bibitem[{{Savage} {et~al.}(2003){Savage}, {Sembach}, {Wakker}, {Richter},
  {Meade}, {Jenkins}, {Shull}, {Moos}, \& {Sonneborn}}]{Savage2003}
{Savage}, B.~D., {Sembach}, K.~R., {Wakker}, B.~P., {et~al.} 2003, \apjs, 146,
  125

\bibitem[{{Sembach} {et~al.}(1995){Sembach}, {Savage}, {Lu}, \&
  {Murphy}}]{Sembach1995}
{Sembach}, K.~R., {Savage}, B.~D., {Lu}, L., \& {Murphy}, E.~M. 1995, \apj,
  451, 616

\bibitem[{{Sembach} {et~al.}(1999){Sembach}, {Savage}, {Lu}, \&
  {Murphy}}]{Sembach1999}
{Sembach}, K.~R., {Savage}, B.~D., {Lu}, L., \& {Murphy}, E.~M. 1999, \apj,
  515, 108

\bibitem[{{Sembach} {et~al.}(2003){Sembach}, {Wakker}, {Savage}, {Richter},
  {Meade}, {Shull}, {Jenkins}, {Sonneborn}, \& {Moos}}]{Sembach2003}
{Sembach}, K.~R., {Wakker}, B.~P., {Savage}, B.~D., {et~al.} 2003, \apjs, 146,
  165

\bibitem[{{Shapiro} \& {Field}(1976)}]{Shapiro1976}
{Shapiro}, P.~R. \& {Field}, G.~B. 1976, \apj, 205, 762

\bibitem[{{Shull} {et~al.}(2009){Shull}, {Jones}, {Danforth}, \&
  {Collins}}]{Shull2009}
{Shull}, J.~M., {Jones}, J.~R., {Danforth}, C.~W., \& {Collins}, J.~A. 2009,
  \apj, 699, 754

\bibitem[{{Wakker}(2006)}]{Wakker2006}
{Wakker}, B.~P. 2006, \apjs, 163, 282

\bibitem[{{Wakker} \& {van Woerden}(1997)}]{Wakker1997}
{Wakker}, B.~P. \& {van Woerden}, H. 1997, \araa, 35, 217

\bibitem[{{Wang}(2009)}]{Wang2009}
{Wang}, Q.~D. 2009, in American Institute of Physics Conference Series, Vol.
  1156, American Institute of Physics Conference Series, ed. {R.~K.~Smith,
  S.~L.~Snowden, \& K.~D.~Kuntz}, 257--267

\bibitem[{{Wilms} {et~al.}(2000){Wilms}, {Allen}, \& {McCray}}]{Wilms}
{Wilms}, J., {Allen}, A., \& {McCray}, R. 2000, \apj, 542, 914

\bibitem[{{Yamagata} \& {Yoshii}(1994)}]{Yamagata1994}
{Yamagata}, T. \& {Yoshii}, Y. 1994, in IAU Symposium, Vol. 161, Astronomy from
  Wide-Field Imaging, ed. {H.~T.~MacGillivray}, 420

\bibitem[{{Yao} \& {Wang}(2005)}]{Yao2005}
{Yao}, Y. \& {Wang}, Q.~D. 2005, \apj, 624, 751

\bibitem[{{Yao} {et~al.}(2009){Yao}, {Wang}, {Hagihara}, {Mitsuda}, {McCammon},
  \& {Yamasaki}}]{Yao09b}
{Yao}, Y., {Wang}, Q.~D., {Hagihara}, T., {et~al.} 2009, \apj, 690, 143

\bibitem[{{York} {et~al.}(1982){York}, {Songaila}, {Blades}, {Cowie}, {Morton},
  \& {Wu}}]{York1982}
{York}, D.~G., {Songaila}, A., {Blades}, J.~C., {et~al.} 1982, \apj, 255, 467

\end{thebibliography}

\end{document}